\documentclass[aps,prd,nofootinbib,showkeys,preprint,floatfix]{revtex4}

\usepackage{amsmath}
\usepackage{amssymb}
\usepackage{graphicx}
\usepackage{rotating}
\usepackage{color} 
\usepackage{multirow} 
 \usepackage[T1]{fontenc}

\def\vev#1{\left\langle #1\right\rangle}

\newcommand{\AddrAHEP}{
  {\it AHEP Group, Instituto de F\'{\i}sica Corpuscular --
    C.S.I.C./Universitat de Val{\`e}ncia \\
    Edificio de Institutos de Paterna, Apartado 22085,
  E--46071 Val{\`e}ncia, Spain}}

\newcommand{\AddrLisb}{%
 Departamento de F\'\i sica and CFTP, Instituto Superior T\'ecnico\\
          Av. Rovisco Pais 1, 1049-001 Lisboa, Portugal }

\newcommand{\AddrWur}{%
Institut f\"ur Theoretische Physik und Astronomie, 
Universit\"at W\"urzburg\\
Am Hubland, 
97074 Wuerzburg}

\def\gsim{\raise0.3ex\hbox{$\;>$\kern-0.75em\raise-1.1ex\hbox{$\sim\;$}}}
\def\lsim{\raise0.3ex\hbox{$\;<$\kern-0.75em\raise-1.1ex\hbox{$\sim\;$}}}

\begin{document}

\preprint{CFTP/09-16}  
\preprint{IFIC/09-08}  


\title{Flavour violation at the LHC: type-I versus type-II
 seesaw in minimal supergravity}


\author{J. N. Esteves}\email{joaomest@cftp.ist.utl.pt}\affiliation{\AddrLisb}
\author{M.~Hirsch} \email{mahirsch@ific.uv.es}\affiliation{\AddrAHEP}
\author{W. Porod} \email{porod@physik.uni-wuerzburg.de}\affiliation{\AddrWur}
\author{J.~C.~Romao}\email{jorge.romao@ist.utl.pt}\affiliation{\AddrLisb}
\author{J.~W.~F.~Valle} \email{valle@ific.uv.es}\affiliation{\AddrAHEP} 
\author{A.~Villanova del Moral}
\email{albert@cftp.ist.utl.pt}\affiliation{\AddrLisb}



\begin{abstract}

    We reconsider the role that the possible detection of lepton
    flavour violating (LFV) decays of supersymmetric particles at the
    Large Hadron Collider (LHC) can play in helping reconstruct the
    underlying neutrino mass generation mechanism within the simplest
    high-scale minimal supergravity (mSUGRA) seesaw schemes.
    We study in detail the LFV scalar tau decays at the LHC, assuming
    that the observed neutrino masses arise either through the pure
    type-I or the simpler type-II seesaw mechanism.
    We perform a scan over the mSUGRA parameter space in order to
    identify regions where lepton flavour violating decays of
    $\chi^0_2$ can be maximized, while respecting current low-energy
    constraints, such as those coming from the bounds on Br($\mu\to e
    \gamma$).  We estimate the cross section for 
    $\chi^0_2\to \chi^0_1 + \tau + \mu $. 
    Though insufficient for a full reconstruction of the seesaw, the
    search for LFV decays of supersymmetric states at the LHC 
    brings complementary information to that coming from low energy
    neutrino oscillation experiments and LFV searches.

\end{abstract}

\maketitle

\section{Introduction}
\label{sec:int}

Neutrino oscillation experiments \cite{Fukuda:1998mi} have provided
the first signal of physics beyond the standard model (SM). These
measurements show that (a) neutrinos have a non-zero mass and (b) {\em
  lepton flavour} is violated. So far there is no experimental data
that indicates that {\em lepton number} is also broken. However,
  one expects neutrinos to be {\em Majorana} particles, their mass at
  low energy being described by a unique ($\Delta L=2$) dimension-5
  operator \cite{Weinberg:1979sa}
\begin{equation}\label{eq:dim5}
\mathcal{L}_{m_{\nu}} = \frac{f}{\Lambda} (H L) (H L).
\end{equation}
where $\Lambda$ is some high energy scale, $f$ a dimensionless
coupling constant, and $H$ ($L$) the Higgs boson (lepton) doublets,
respectively.  Many model realizations of eq. (\ref{eq:dim5}) exist,
the most famous being the seesaw mechanism. The latter can be
implemented via the exchange of a heavy singlet fermion, usually
called type-I
seesaw~\cite{Minkowski:1977sc,seesaw,MohSen,Schechter:1980gr}.  The
exchange of a scalar triplet \cite{Konetschny:1977bn,Marshak:1980yc,
Schechter:1980gr,Cheng:1980qt,Lazarides:1980nt,Mohapatra:1980yp} is 
now known as type-II seesaw~\footnote{This is the opposite convention
  to that used in \cite{Schechter:1980gr}.}. The exchange of a
fermionic triplet \cite{Foot:1988aq} is called type-III seesaw in
\cite{Ma:1998dn}.  A list of generic 1-loop realizations of
eq. (\ref{eq:dim5}) can also be found in \cite{Ma:1998dn}. Further
seesaw realizations, such as  the inverse and the linear
seesaw, are discussed in \cite{Valle:2006vb}.

At ``low'' energies one cannot decide whether tree-level or loop
physics generates eq.~(\ref{eq:dim5}), nor can any measurements of
neutrino angles, phases or masses distinguish between the
different tree-level seesaw realizations.
Under the assumption of a pure type-I or pure type-II minimal
supergravity seesaw mechanisms, we reconsider here the prospects for
reconstructing the underlying high energy parameters from a
combination of different measurements.
Clearly, observables outside the neutrino sector are needed in order
to ultimately learn about the high energy parameters characterizing
the seesaw. 
The classical tree-level realizations of the simplest
type-I seesaw mechanism, unfortunately, can not be put to the test in
a direct way. This can be straightforwardly understood by inverting
eq.~(\ref{eq:dim5}), which results in $\Lambda \sim f \Big(\frac{\rm
  0.05 \hskip1mm eV}{m_{\nu}}\Big) 10^{15}$ GeV.

If the CERN LHC, due to take first data, finds signs of
electroweak scale supersymmetry, indirect insight into the high-energy
world might become possible through the search for flavour 
violation effects~\cite{delAguila:2008iz,Raidal:2008jk}. 
Starting from flavour
diagonal soft supersymmetry breaking terms at some high energy
``unification'' scale, flavour violation appears at lower energies due
to the renormalization group evolution of the soft breaking parameters
\cite{Hall:1985dx,Donoghue:1983mx}.  
If the seesaw mechanism is responsible for the
observed neutrino masses, the neutrino Yukawa couplings leave their
imprint in the slepton mass matrices as first shown in
\cite{Borzumati:1986qx}. Potentially large LFV is then induced by the
flavour off-diagonal structure in the Yukawa couplings required by the
large mixing angles observed in oscillation experiments
\cite{Maltoni:2004ei}. Expectations for LFV decays such as $l_i \to
l_j +\gamma$ and $l_i \to 3 l_j$ in the supersymmetric seesaw have
been studied in
\cite{Hisano:1995nq,Hisano:1995cp,Deppisch:2002vz,Arganda:2005ji,%
  Antusch:2006vw,Deppisch:2004fa}. For the related process of $\mu-e$
conversion in nuclei see, for example
\cite{Arganda:2007jw,Deppisch:2005zm}.  The potential of LHC
experiments in probing the allowed seesaw parameters through
measurements of masses and branching ratios of supersymmetric
particles has been also discussed in
Refs.~\cite{Hisano:1998wn,Blair:2002pg,%
  Freitas:2005et,Buckley:2006nv,Deppisch:2007xu}.

In two previous studies \cite{Hirsch:2008dy,Hirsch:2008gh} we have
pointed out that ratios of branching ratios are especially useful for
learning about the unknown seesaw parameters. In \cite{Hirsch:2008dy}
the case of type-I seesaw was discussed, whereas \cite{Hirsch:2008gh}
addresses the case of seesaw type-II. For the type-I seesaw, there are
in general too many unknown parameters that preclude making any
definite predictions for LFV decays. In contrast, in the simplest
type-II seesaw model (with only one triplet coupling to standard model
leptons) neutrino mixing angles can be related to ratios such as
Br(${\tilde\tau}_2\to e\chi^0_1$)/Br(${\tilde\tau}_2\to \mu\chi^0_1$).

It has been shown that, to a good approximation, such ratios do not
depend on the mSUGRA parameter values. However, from an experimental
point of view, calculations of absolute event rates are needed, before
ratios of different final state channels can be studied. In
\cite{Hirsch:2008dy,Hirsch:2008gh} we took as reference just a few
benchmark mSUGRA points, for which we have made detailed studies.
In the present paper we calculate branching ratios and event rates
over a large region of mSUGRA parameter space, in order to identify
the maximal number of events one can expect in experiments at the LHC,
while still respecting all low-energy constraints.

The rest of the paper is organized as follows. In the next section we
give a short summary of the theoretical setup.  In section III we
describe our numerical procedure and present our results. Finally we
close in section IV with a discussion and a short summary.

\section{Theory setup}
\label{sec:theory-setup}

In order to fix the notation, we will briefly discuss the main
features of the seesaw mechanism and mSUGRA.  The type-I
supersymmetric seesaw consists in extending the particle content of
the MSSM by three gauge singlet ``right-handed'' neutrino
superfields. The leptonic part of the superpotential is then
\begin{equation}\label{su_pot}
W  =  Y_{e}^{ji}{\widehat L}_i{\widehat H_d}{\widehat E^c}_j
  + Y_{\nu}^{ji}{\widehat L}_i{\widehat H_u}{\widehat N^c}_j
  + M_{i}{\widehat N^c}_i{\widehat N^c}_i, 
 \quad i,j=1,\ldots,3 ,
\end{equation}
where $Y_e$ and $Y_{\nu}$ denote the charged lepton and neutrino
Yukawa couplings, while ${\widehat N^c}_i$ are the ``right-handed''
neutrino superfields with $M_{i}$ Majorana mass terms. 
One can always choose a basis in
which the Majorana mass matrix of the ``right-handed'' neutrinos is
brought to diagonal form $ \hat M_R=\mathrm{diag}(M_1,M_2,M_3)$.
Without loss of generality we will also assume that eq. (\ref{su_pot})
is written in the basis where the charged lepton Yukawa matrix is
already diagonal. In this simple setup, the type-I seesaw model, as
defined by eq. (\ref{su_pot}), is characterized by a total of 21
parameters, from which only 12 are measurable in the low-energy
theory, as we discuss below.

The effective mass matrix of the ``left-handed'' neutrinos at low
energies is then given as
\begin{equation}\label{meff}
m_{\nu} = - \frac{v_u^2}{2} Y_{\nu}^T\cdot {\hat M}_{R}^{-1}\cdot Y_{\nu},
\end{equation}
so that, for each ``right-handed'' neutrino, there is one non-zero
eigenvalue in $m_{\nu}$.  In eq.~(\ref{meff}) we use the notation
$\vev{H_{u,d}}=\frac{v_{u,d}}{\sqrt{2}}$ for the vacuum expectation
values of the neutral components of the Higgs boson doublets.

The parameters of eq.(\ref{su_pot}) are defined at the Grand Unified
Theory (GUT) scale, whereas the entries of eq. (\ref{meff}) are
measured at low energies.
In order to connect these two scales we numerically solve the full set
of renormalization group equations (RGE)
\cite{Hisano:1995cp,Antusch:2002ek}.

Being complex symmetric, the light Majorana neutrino mass matrix
  in eq.~(\ref{meff}), is diagonalized by a unitary $3\times 3$ matrix
  $U$~\cite{Schechter:1980gr}
\begin{equation}\label{diagmeff}
{\hat m_{\nu}} = U^T \cdot m_{\nu} \cdot U\ .
\end{equation}

Inverting the seesaw equation, eq.~(\ref{meff}), allows to express 
$Y_{\nu}$ as \cite{Casas:2001sr}
\begin{equation}\label{Ynu}
Y_{\nu} =\sqrt{2}\frac{i}{v_u}\sqrt{\hat M_R}\cdot R \cdot \sqrt{{\hat
    m_{\nu}}} \cdot U^{\dagger},
\end{equation}
where $\hat m_{\nu}$ is the diagonal matrix with $m_i$ eigenvalues and
in general $R$ is a complex orthogonal matrix. Note that, in the
special case $R=1$, $Y_{\nu}$ contains only ``diagonal'' products
$\sqrt{M_im_{i}}$. In this simplified case the 18 parameters in
$Y_{\nu}$ are reduced to 12. Note that in general type-I seesaw
schemes, the unitary matrix diagonalizing the effective neutrino
mass matrix differs from the lepton mixing matrix by terms of order
$D/M_R$, where the $D = Y_\nu v_u$. For the high-scale schemes
considered here one can safely neglect these
deviations~\footnote{However for other type-I schemes, like the
  inverse seesaw~\cite{Deppisch:2004fa,Deppisch:2005zm} this
  approximation fails and leads to large LFV from right-handed
  neutrino exchange, even in the absence of supersymmetric
  contributions. For a systematic perturbative seesaw
  diagonalization method that covers all cases see the second paper
  in Ref.~\cite{Schechter:1980gr}.}.  In this case we can set the
diagonalization matrix as the lepton mixing matrix (partially)
determined in neutrino oscillation measurements. 

Implementing the type-II seesaw mechanism within supersymmetry
requires at least two $SU(2)$ triplet states $T_{1,2}$. A scalar
triplet with mass below the GUT scale changes the running of $g_1$ and
$g_2$ in an unwanted way and gauge coupling unification is lost.  If
instead one adds only complete $SU(5)$ multiplets (or GUT multiplets
which can be decomposed into complete $SU(5)$ multiplets) to the
standard model particle content, the scale where couplings unify
remains the same (at one loop level), only the value of the GUT
coupling itself changes \cite{Langacker:1980js}.

Our numerical calculation uses an $SU(5)$ inspired model
\cite{Rossi:2002zb}, which adds a pair of ${\bf 15}$ and
$\overline{\bf 15}$ to the Minimal Supersymmetric Standard Model
(MSSM) particle spectrum. This variant of the type-II seesaw
mechanism, as discussed above, allows us to maintain gauge coupling
unification even for $M_T\ll M_G$, $M_G$ being the unification 
scale. Under $SU(3)\times SU_L(2) \times U_Y(1)$ the ${\bf 15}$ 
decomposes as
\begin{eqnarray}\label{eq:15}
{\bf 15} & = &  S + T + Z \\ \nonumber
S & \sim  & (6,1,-\frac{2}{3}), \hskip10mm
T \sim (1,3,1), \hskip10mm
Z \sim (3,2,\frac{1}{6}).
\end{eqnarray}
$T$ has the correct quantum numbers to generate the dimension-5
operator of eq. (\ref{eq:dim5}). The $SU(5)$ invariant superpotential
reads 
\begin{eqnarray}\label{eq:pot15}
W & = & \frac{1}{\sqrt{2}}{\bf Y}^{15}_{ij}\, {\bf \bar 5}_i \cdot {\bf 15} 
\cdot {\bf \bar 5}_j 
+ \frac{1}{\sqrt{2}}\lambda_1 {\bf \bar 5}_H \cdot {\bf 15} \cdot 
{\bf \bar 5}_H 
+ \frac{1}{\sqrt{2}}\lambda_2 {\bf 5}_H \cdot \overline{\bf 15} \cdot
{\bf 5}_H 
+ {\bf Y}^5_{ij}\, {\bf 10}_i \cdot {\bf \bar 5}_j \cdot {\bf \bar 5}_H  \nonumber\\
 & + & {\bf Y}^{10}_{ij}\, {\bf 10}_i \cdot {\bf 10}_j 
\cdot {\bf 5}_H + M_{15} {\bf 15} \cdot \overline{\bf 15} 
+ M_5 {\bf \bar 5}_H \cdot {\bf 5}_H \ .
\end{eqnarray}
Here, ${\bf \bar 5}=(d^c,L)$, ${\bf 10}=(u^c,e^c,Q)$, ${\bf 5}_H =(t,H_2)$ and 
${\bf \bar 5}_H=({\bar t},H_1)$. Below the GUT scale, in the $SU(5)$-broken 
phase, the superpotential contains the terms 
\begin{eqnarray}\label{eq:broken}
 &  & \frac{1}{\sqrt{2}}(Y_{T}^{ij}\, L_i T_1 L_j +  Y_{S}^{ij}\, d_i^c S d_j^c) 
+ Y_{Z}^{ij}\, d^c_i Z L_j + Y_{d}^{ij}\,  d^c_i Q_j H_d + Y_{u}^{ij}\, u_i^c Q_j H_u  
+ Y_{e}^{ij}\, e^c_i L_j H_d \nonumber \\ 
& & +\frac{1}{\sqrt{2}}(\lambda_1 H_d T_1 H_d +\lambda_2 H_u T_2 H_u) 
+ M_T T_1 T_2 + M_Z Z_1 Z_2 + M_S S_1 S_2 + \mu H_d H_u \ . 
\end{eqnarray}
As long as $M_Z \sim M_S \sim M_T$, gauge coupling
unification will be preserved. Note that exact equality is not
required for a successful unification.  In our numerical studies we
have taken into account the different running of these mass
parameters. 

Integrating out the heavy triplets at their mass scale, the dimension-5
operator of eq.~(\ref{eq:dim5}) is generated and after electroweak
symmetry breaking the resulting neutrino mass
matrix can be written as
\begin{eqnarray}\label{eq:ssII}
m_\nu=\frac{v_u^2}{2} \frac{\lambda_2}{M_T}Y_T \ .
\end{eqnarray}
As in the case of the type-I seesaw, eq. (\ref{eq:ssII}) depends on
the energy scale. In order to compute the neutrino mass $m_\nu$
measured at low energies, one needs to know $\lambda_2$, $Y_T$ and
$M_T$ as input parameters at the high energy scale. As will be
discussed in section \ref{sec:num}, one can use an iterative procedure
in order to find the high scale parameters from the low energy
measured quantities.

Note that, without loss of generality, we have the freedom to
write eqs.~(\ref{su_pot}) and (\ref{eq:pot15}) in the basis where
the charged lepton mass matrix is diagonal, fitting the
corresponding Yukawa couplings so as to reproduce the three measured
charged lepton masses.
However note that there are important differences between the type-I
and type-II seesaw schemes. For example, in contrast to type-I, in a
pure type-II seesaw scheme the unitary matrix $U$ that diagonalizes
eq.~(\ref{eq:ssII}) coincides with the lepton mixing matrix studied
in neutrino oscillations.
Moreover, in sequential type-I seesaw for each ``right-handed''
neutrino added there is one non-zero light neutrino mass
eigenstate~\footnote{We do not consider here the possibility of 
  having just two right-handed neutrino states in the type-I seesaw, 
  called (3,2) in Ref.~\cite{Schechter:1980gr}. This could well
  account for the current neutrino data with just 12 parameters, 
  instead of the 18 characterizing the sequential (3,3) seesaw 
  considered here.}. 
In contrast, in type-II seesaw one can produce three neutrino masses
with just one pair of triplet superfields, with only one triplet
directly coupling to leptons.
This implies that in the minimal type-II seesaw one has less
parameters than in the sequential type-I seesaw. Indeed from the 12
parameters in the complex symmetric $Y_T$ matrix, one can remove 3
phases by redefining the charged leptons~\cite{Schechter:1980gr}. In
addition, from the 3 complex parameters $\lambda_{1,2}$ and $M_T$,
one does not enter, as only one of the triplets couples to leptons,
and finally, two of the three phases can also be removed by field
redefinitions. The net result is that there are only 11 physical
parameters governing neutrino physics~\cite{Hirsch:2008gh}.
This number is substantially smaller than the 18 free parameters
describing the simplest type-I seesaw scheme containing three
``right-handed'' neutrinos~\cite{Santamaria:1993ah}~\footnote{We
  are treating the three charged lepton masses as experimentally
  determined parameters.}.

At low energies a maximum of 9 neutrino parameters can be fixed by
measuring lepton properties: 3 neutrino masses, 3 mixing angles and 3
CP phases. Thus from neutrino data only, neither type-I nor type-II
seesaw schemes can be completely reconstructed, even in their simplest
realizations. However, especially important in the following is the
fact that low-energy neutrino angles are directly related to the
high-energy Yukawa matrix in the type-II seesaw, whereas no
such simple connection exists in the seesaw type-I (see also the
discussion in \cite{Ellis:2002fe}).

As already commented, to a good approximation the
  lepton mixing matrix may be taken in unitary form, with three mixing
  angles $\theta_{ij}$, and three physical CP phases
  $\phi_{ij}$~\cite{Schechter:1980gr}. Of these only the leptonic
  analogue of the Kobayashi-Maskawa phase $\delta$, taken as the
  invariant combination $\delta \equiv \phi_{12} - \phi_{13}
  +\phi_{23} $ would enter the class of LFV processes discussed in this
  paper, so that we get the standard form,
\begin{eqnarray}\label{def:unu}
U=
\left(
\begin{array}{ccc}
 c_{12}c_{13} & s_{12}c_{13}  & s_{13}e^{-i\delta}  \\
-s_{12}c_{23}-c_{12}s_{23}s_{13}e^{i\delta}  & 
c_{12}c_{23}-s_{12}s_{23}s_{13}e^{i\delta}  & s_{23}c_{13}  \\
s_{12}s_{23}-c_{12}c_{23}s_{13}e^{i\delta}  & 
-c_{12}s_{23}-s_{12}c_{23}s_{13}e^{i\delta}  & c_{23}c_{13}  
\end{array}
\right) 
\end{eqnarray}
where $s_{ij}\equiv \sin\theta_{ij}$, $c_{ij}=\cos\theta_{ij}$.
Since no current experiment is sensitive enough to probe leptonic CP
violation we take, for simplicity, $\delta=0$.
Neutrino oscillation experiments can be fitted with either a normal
hierarchical spectrum (NH), or with an inverted hierarchy (IH) one. If
one does not insist in ordering the neutrino mass eigenstates
$m_{\nu_i}$, $i=1,2,3$ with respect to increasing mass, the matrix $U$
can describe both possibilities without re-ordering of angles.  In
this convention, which we will use in the following, $m_{\nu_1} \simeq
0$ ($m_{\nu_3}\simeq 0$) corresponds to normal (inverse) hierarchy and
$s_{12}$, $s_{13}$ and $s_{23}$ are the angles in both types of
spectra. Basically $s_{12}$ is measured in solar + reactor
experiments, $s_{23}$ in atmospheric + accelerator experiments and
$s_{13}$ is constrained by reactor neutrino oscillation data.

In the general MSSM, LFV off-diagonal entries in the slepton mass
matrices involve additional free parameters which arise from the
mechanism of supersymmetry breaking. In order to relate LFV in the
slepton sector with the LFV encoded in $Y_{\nu}$ or $Y_T$ one must
assume some particular scheme for supersymmetry breaking. For
simplicity and definiteness we will adopt mSUGRA boundary conditions,
characterized by four continuous real and one discrete free parameter,
usually denoted as
\begin{equation}\label{sugra-par}
m_0, \hskip2mm M_{1/2}, \hskip2mm A_0, \hskip2mm \tan\beta, \hskip2mm 
{\rm Sgn}(\mu)\ .
\end{equation}
Here, $m_0$ is the common scalar mass, $M_{1/2}$ the gaugino mass 
and $A_0$ the common trilinear parameter, all defined at the 
grand unification scale, $M_{G} \simeq 2 \cdot 10^{16}$ GeV. 
The remaining two parameters are $\tan\beta = v_u/v_d$ and the 
sign of the Higgs mixing parameter $\mu$. 

In order to have a qualitative understanding of the magnitudes
of the LFV rates we first present approximate leading-log analytical
solutions for the renormalization group equations~\footnote{Note
  that in the numerical code that leads to the results presented in
  our plots we have numerically solved the full set of RGEs.}. For
the case of type-I seesaw, the LFV elements induced in the charged
left-slepton mass matrix by renormalization group evolution can be
approximated as~\cite{Hisano:1995cp}
\begin{equation}\label{running-I}
(\Delta M_{\tilde L}^2)_{ij} =
 -\frac{1}{8\pi^2}(3 m_0^2 + A_0^2) 
  (Y_{\nu}^{\dagger}LY_{\nu})_{ij}\,, 
\end{equation}
where $Y_{\nu}$ is given in terms of the neutrino parameters by
eq.~(\ref{Ynu}) and the factor $L$ is defined as
\begin{equation}\label{deffacL}
L_{kl} = \log\Big(\frac{M_G}{M_{k}}\Big)\delta_{kl}\,. 
\end{equation}

Similarly, one can get an analogous approximate expression for the
off-diagonal elements of the charged left-slepton mass matrix
characterizing LFV in type-II seesaw schemes~\cite{Rossi:2002zb}.

\section{Numerical results}
\label{sec:num}

Due to the non-trivial structure of the neutrino Yukawa matrix
$Y_{\nu}$ in eq.~(\ref{Ynu}) and of $Y_T$ in eq.~(\ref{eq:ssII})
for type-I and type-II seesaw, respectively, the slepton mass
matrices contain calculable LFV
entries~\cite{Hall:1985dx,Borzumati:1986qx}.
In order to determine their magnitude we solve the complete set of
renormalization group equations, given
in~\cite{Hisano:1995cp,Antusch:2002ek,Rossi:2002zb}.
All results presented below have been obtained with the lepton flavour
violating version of the program package SPheno~\cite{Porod:2003um},
where the RGEs for the MSSM part are implemented at the 2-loop level.
For definiteness we set neutrino mass squared differences to
their current best fit values~\cite{Maltoni:2004ei} and fix the
angles to the Tri-Bi-Maximal (TBM) values~\cite{Harrison:2002er}.

Fixing the values of other mSUGRA parameters, we used SPheno to
perform a numerical scan over the $m_0$-$M_{1/2}$ plane. For each
point in this plane, we adjust the value of $M_R$ ($M_T$) in order
to keep the low energy LFV observable BR($\mu\to e \gamma$) within
its present experimental upper bound or within the expected
sensitivity of the upcoming experiments~\cite{MEG}.


For type-I seesaw our numerical procedure to fit these masses is as
follows. As we have already commented, the large number of free
parameters characterizing even the simplest type-I seesaw schemes
forces us to make simplifying assumptions in inverting the seesaw
equation, eq.~(\ref{Ynu}). As a first step we assume degenerate
``right-handed'' neutrinos and the simplest possible, flavour-blind,
structure for the matrix $R$, i.e.
\begin{equation}
  \label{eq:R}
  R = 1,~~~~~~~~~~~~~~{\hat M}_{R\, ij} = M_R\ \delta_{ij}\ .
\end{equation}
Moreover, we fix the values of the light neutrino masses and 
Yukawa couplings to reproduce the TBM angle values.  In order to
determine the resulting LFV observables we numerically integrate the
renormalization group equations taking into account the flavour
structure of the $Y_\nu$ matrix. We integrate out every
``right-handed'' neutrino and its superpartner at the scale 
associated to its mass, and calculate the corresponding
contribution to the dimension-five operator which is evolved to 
the electroweak scale. This way we obtain the exact neutrino masses
and mixing angles for this first guess. The difference between the
results  numerically obtained and the input numbers is then
minimized in an iterative procedure until convergence is achieved.


For the type-II seesaw the calculations are performed for the
{\bf 15}-plet case, under the assumption $Y_Z = Y_T = Y_S$ at 
$M_G$, as discussed above, and including the one-loop RGEs for the new
parameters in SPheno. For consistency, we have also included 1-loop
threshold corrections for gauge couplings and gaugino mass parameters
at the scale corresponding to the mass of the triplet, $M_T$. The MSSM
part is implemented at the 2-loop level and, thus, in principle one
should also consistently include the effect of the {\bf 15}-plets for
all parameters at this level. However, as discussed in
~\cite{Hirsch:2008gh}, the correct fit of the neutrino data requires
that either the triplet ({\bf 15}-plet) Yukawa couplings are small
and/or that $M_T$ is close to $M_G$, implying that the ratio $M_T/M_G$
is significantly smaller than $M_G/m_Z$ and thus one expects only
small effects.  Inverting the seesaw equation for any fixed
value of $\lambda_2$ in eq.~(\ref{eq:ssII}), one can get a first
guess of the Yukawa couplings for any fixed values of the light
neutrino masses as a function of the corresponding triplet mass. This
first guess will not give the correct Yukawa couplings, since the
neutrino masses and mixing angles are measured at low energy, whereas
for the calculation of $m_\nu$ we need to insert the parameters at the
high-energy scale. However, we can use this first guess to numerically
run the RGEs to obtain the exact neutrino parameters (at low
energies) for these input values. The difference between the results
obtained numerically and the input can then be minimized in a simple
iterative procedure until convergence is achieved. As long as neutrino
Yukawas are not large, convergence is reached in a few steps.
However, in type-II seesaw schemes, the Yukawa couplings run stronger
than in the type-I seesaw, thus our initial guess can sizeable deviate
from the exact Yukawa coupling values. Since neutrino oscillation data
requires at least one neutrino mass to be larger than about 0.05 eV,
we do not find any solutions for $M_T > 10^{15}$ GeV.

Finally, the calculation of cross sections for the production of
supersymmetric particles was done using Prospino~\cite{prospino}. The
input data was taken from SPheno using the SUSY Les Houches 
Accord standard format~\cite{Skands:2003cj}.

\subsection{Br($\mu\to e +\gamma$) for type-I and type-II} 

\begin{figure}[!hbt]
  \centering
  \begin{tabular}{cc}
    \includegraphics[width=0.48\textwidth]{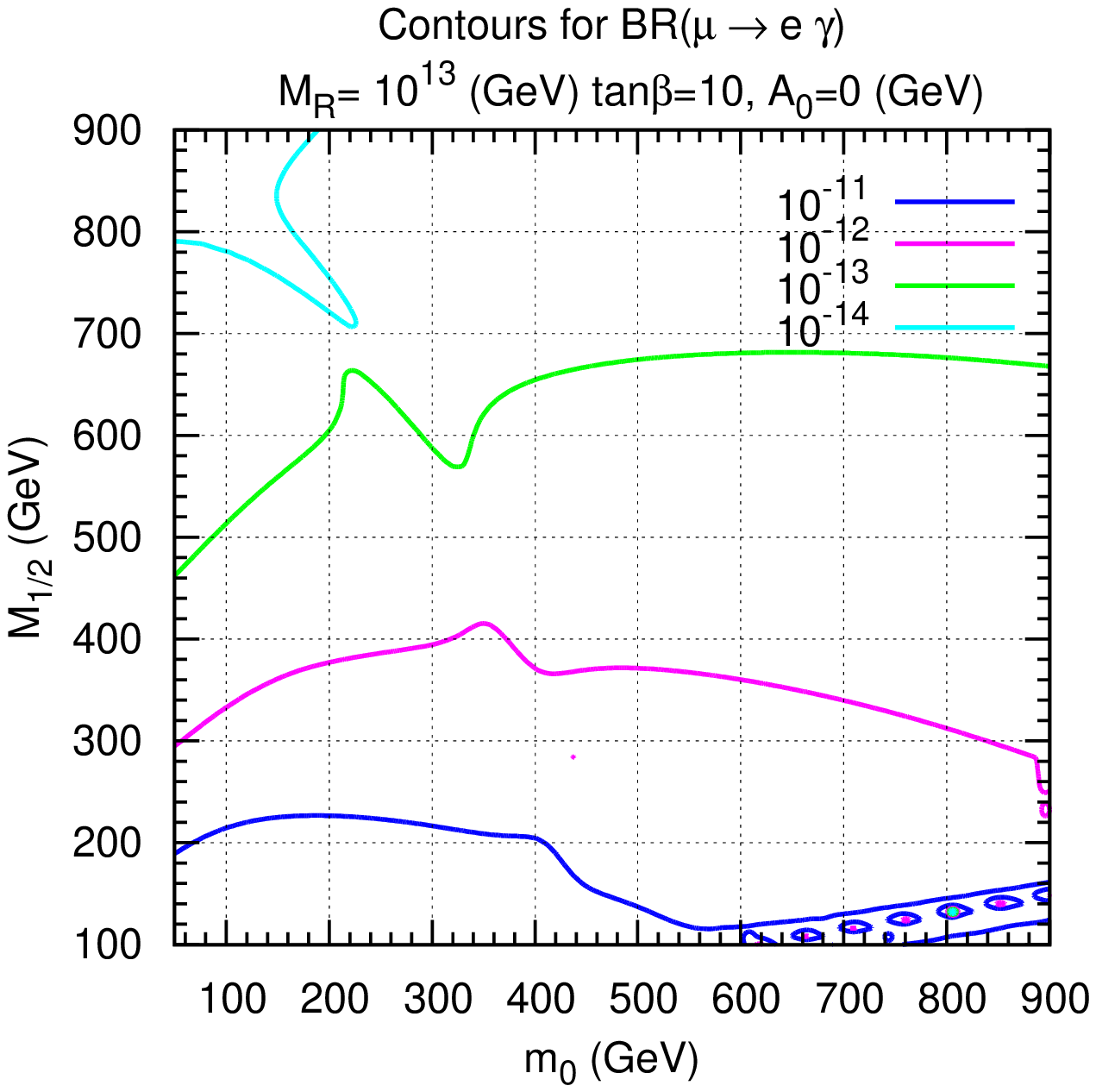}&
    \includegraphics[width=0.48\textwidth]{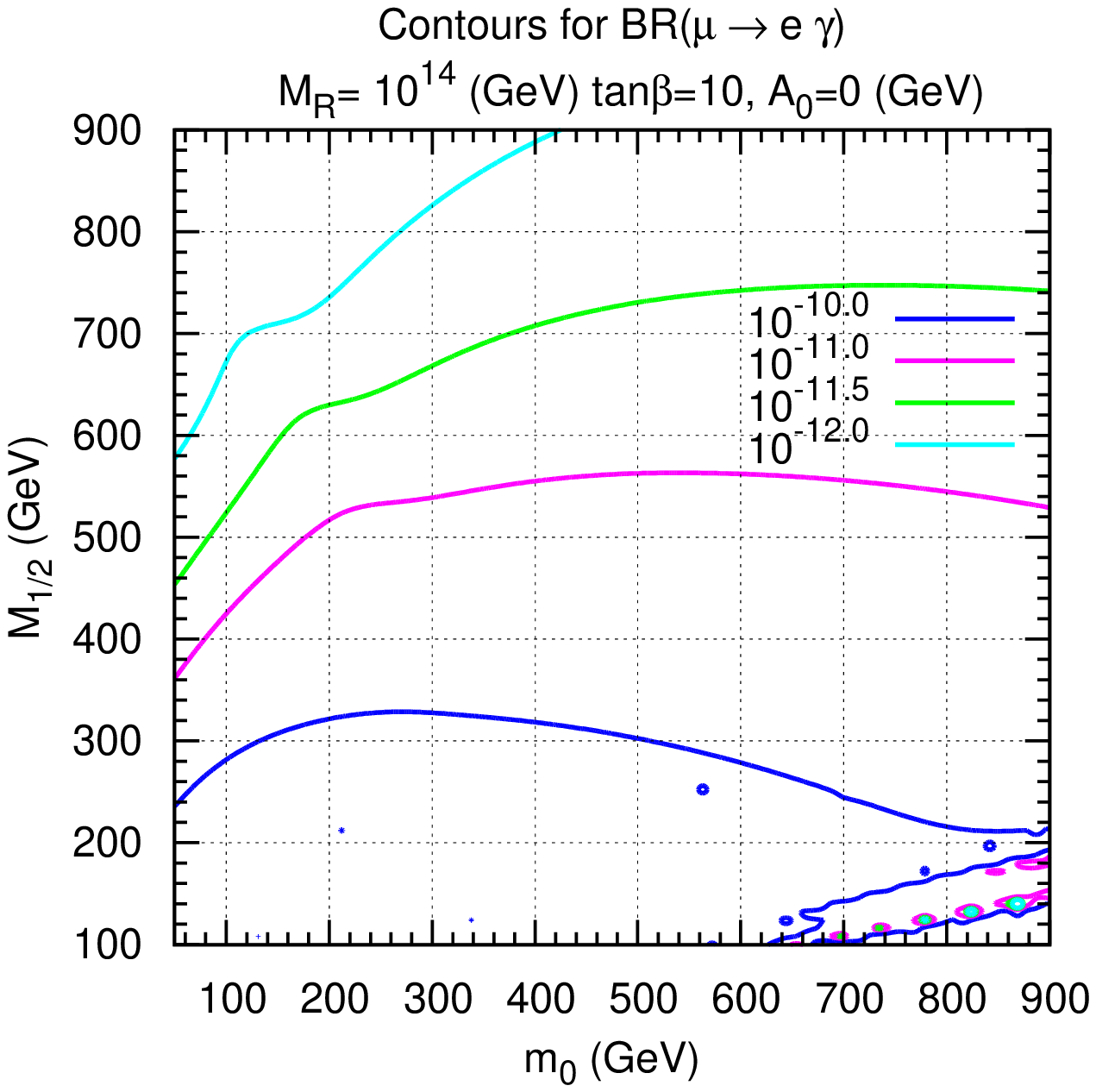}
  \end{tabular}
  \caption{Contours of Br($\mu\to e +\gamma$) in the $m_0,M_{1/2}$
    plane for our standard choice of parameters: $\mu>0$, $\tan\beta=10$
    and $A_0=0$  GeV, 
    for type-I with degenerate RH neutrinos.  On the left panel
    $M_R=10^{13}$ GeV, on the right panel $M_R=10^{14}$ GeV.} 
  \label{fig:1}
\end{figure}

\begin{figure}[!htb]
  \centering
  \begin{tabular}{cc}
    \includegraphics[width=0.48\textwidth]{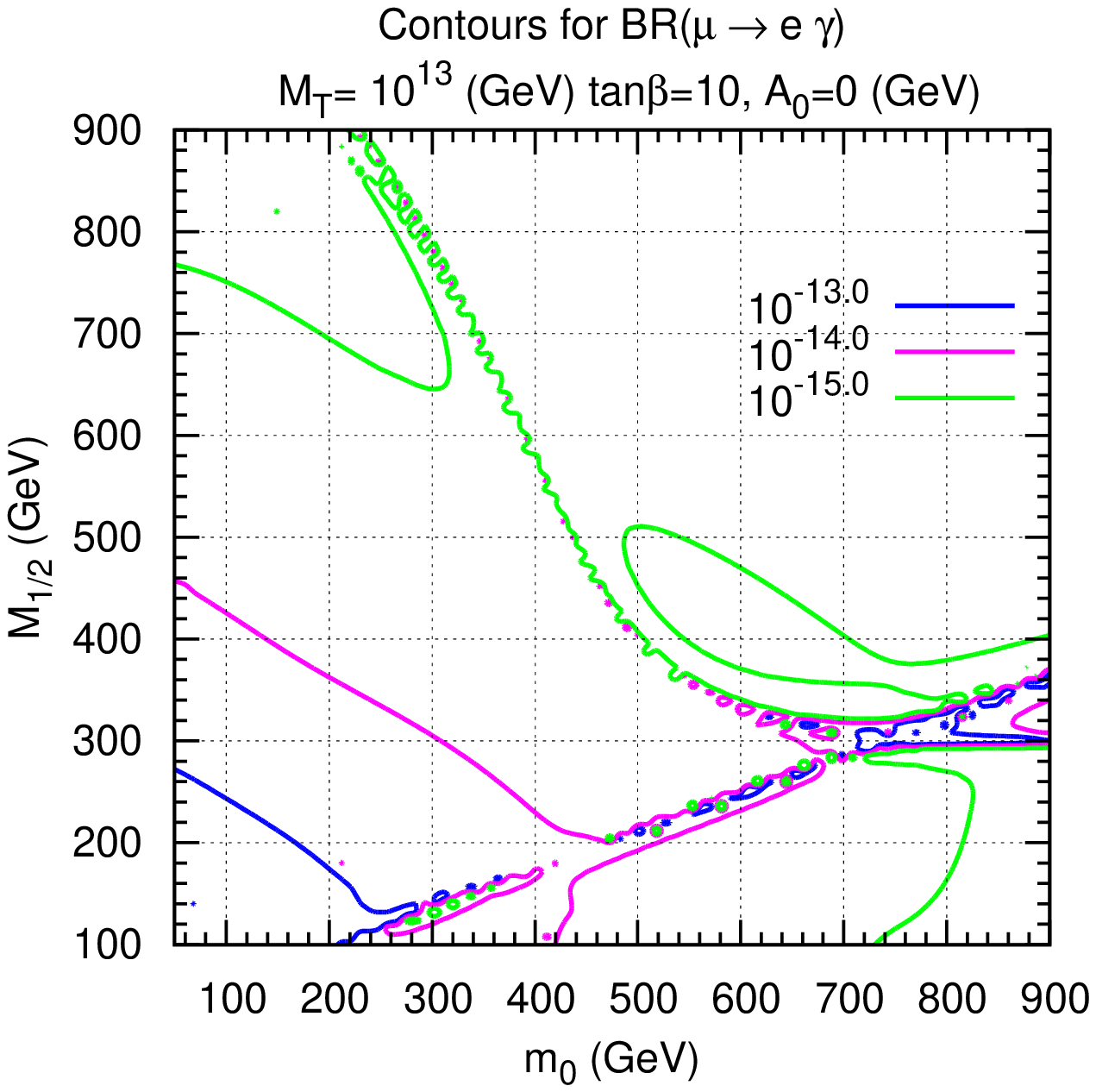}&
    \includegraphics[width=0.48\textwidth]{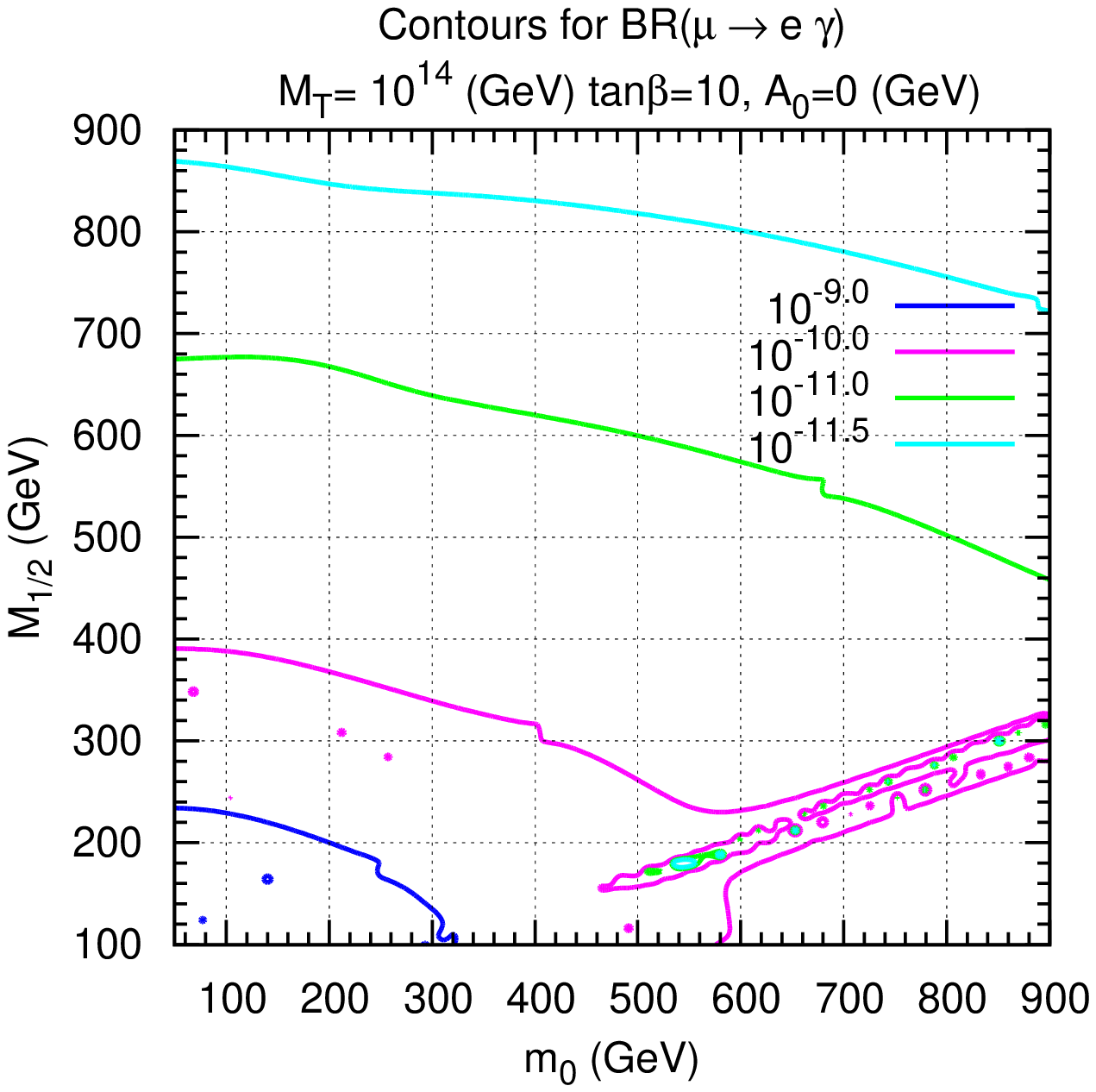}
  \end{tabular}
  \caption{Contours of Br($\mu\to e +\gamma$) in the $m_0,M_{1/2}$ plane 
    for $\lambda_1=0.02$ and $\lambda_2=0.5$ and for our standard choice
    of parameters: $\mu>0$, $\tan\beta=10$ and $A_0=0$ GeV, for
    type-II seesaw.  On the left panel $M_T=10^{13}$ GeV, on the right
    panel $M_T=10^{14}$ GeV.}
  \label{fig:2}
\end{figure}

In Fig.~\ref{fig:1} and Fig.~\ref{fig:2} we show the contours for
Br($\mu\to e +\gamma$) in the $m_0,M_{1/2}$ plane for pure type-I and
type-II seesaw schemes, respectively. On the left panel of
Fig.~\ref{fig:1} we chose a low value for $M_R=10^{13}$, while on the
right panel a value of $10^{14}$ GeV was chosen. In Fig.~\ref{fig:2}
the same dependence is shown for the type-II seesaw mass scale $M_T$.
This shows the dependence of LFV rates on the choice of scale $M_R$
($M_T$). The complicated features displayed on these plots are due to
cancellations between the chargino and neutralino amplitudes
contributing to $\mu\to e +\gamma$, as is well 
known~\cite{Deppisch:2002vz,Hirsch:2008dy}. 
First we note that large parts of parameter space fall within the regions
of sensitivity of upcoming experiments like MEG~\cite{MEG}.
The contours for BR($\mu\to e\,\gamma$) are deformed in type-II as
compared with respect to those for type-I seesaw. The reason for this is
that the addition of gauge non-singlet states in type-II seesaw
increases the dependence on the renormalization scale of the neutrino
Yukawa coupling and also affects the supersymmetric spectrum, which
alters the region where BR($\mu\to e\,\gamma$) is strongly suppressed.
For our subsequent discussion, the most important point is that for
each point in the $m_0,M_{1/2}$ plane there will be a maximum value of
$M_R$ ($M_T$) that will give the maximum possible rates of LFV 
compatible with current experimental bounds, BR($\mu\to e\,\gamma$)
$<1.2 \times 10^{-11}$~\cite{pdg}, or with expected sensitivities to
be reached at upcoming experiments like MEG, BR($\mu\to e\,\gamma$)
$ < 10^{-13}$~\cite{MEG}.

\subsection{LFV stau decays} 
\label{sec:staudecays}

The eagerly awaited production of supersymmetric particles at the
LHC would open new opportunities for the study of flavour violation
in the supersymmetric sector~\cite{delAguila:2008iz}.
Here we study how the LFV decays of staus may provide valuable
cross-checks of neutrino properties determined at low energies as
well as complementary information on the origin of neutrino mass. 
\begin{figure}[!htb]
  \centering
  \begin{tabular}{cc}
    \includegraphics[width=0.48\textwidth]{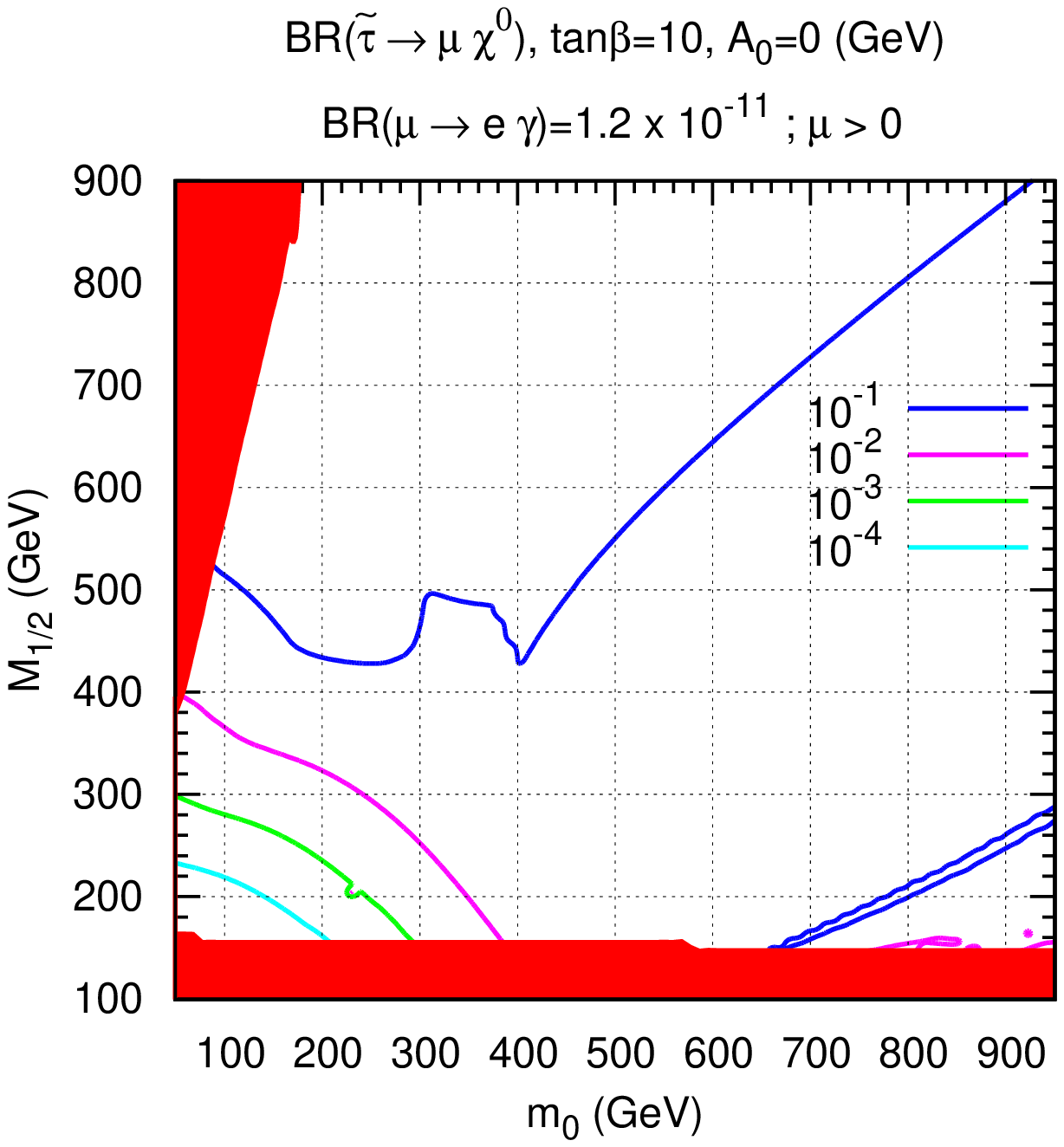}&
    \includegraphics[width=0.48\textwidth]{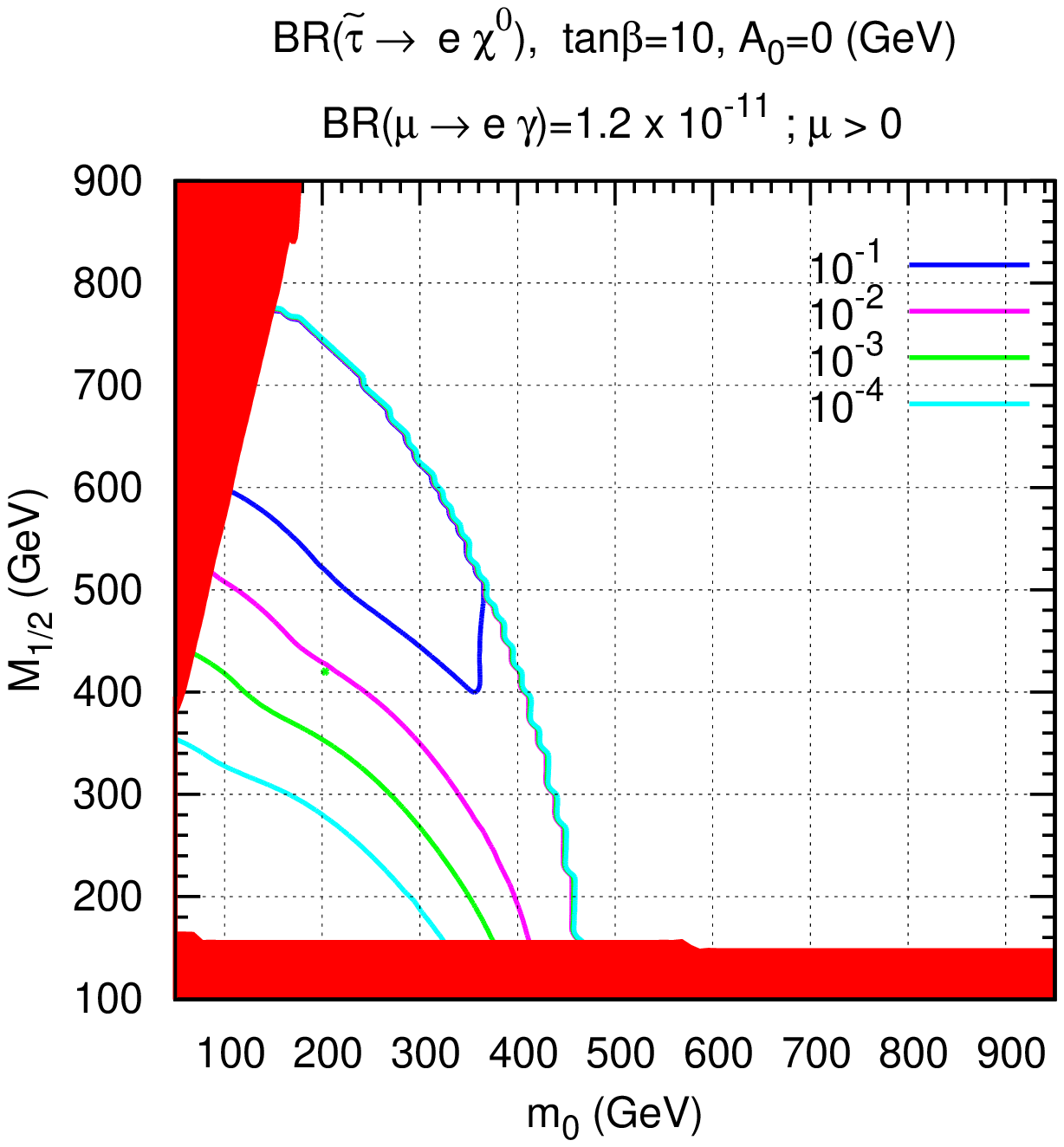}
  \end{tabular}
  \caption{Br(${\tilde\tau}_2 \to \mu +\chi^0_1$) (left panel) and
    Br(${\tilde\tau}_2 \to e +\chi^0_1$) (right panel), in the
    $m_0,M_{1/2}$ plane for our standard choice of parameters:
    $\mu>0$, $\tan\beta=10$ and $A_0=0$ GeV, for type-I seesaw, imposing
    Br($\mu\to e +\gamma) \le 1.2\cdot 10^{-11}$.}
  \label{fig:3}
\end{figure}

\begin{figure}[!htb]
  \centering
  \begin{tabular}{cc}
    \includegraphics[width=0.48\textwidth]{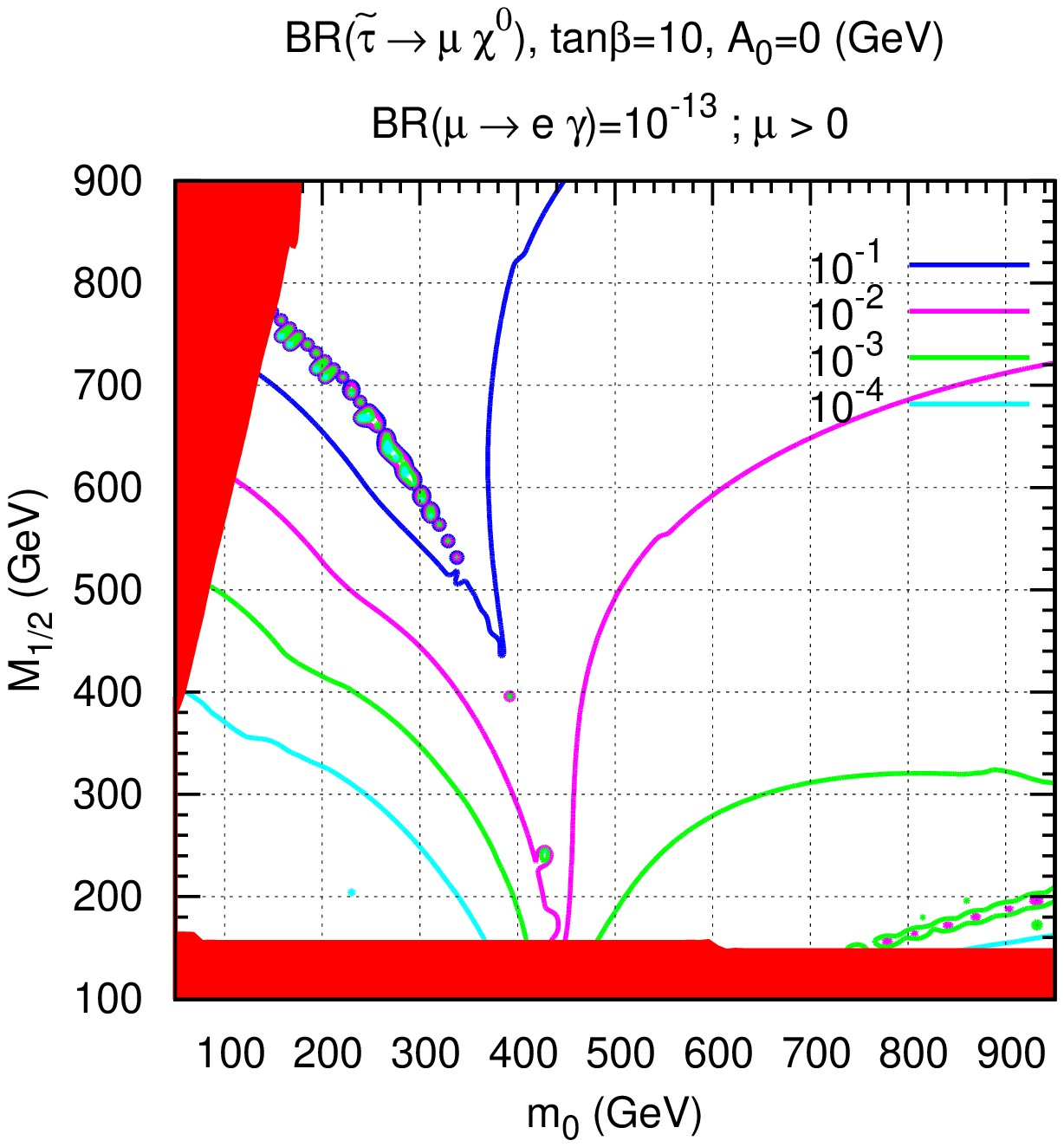}&
    \includegraphics[width=0.48\textwidth]{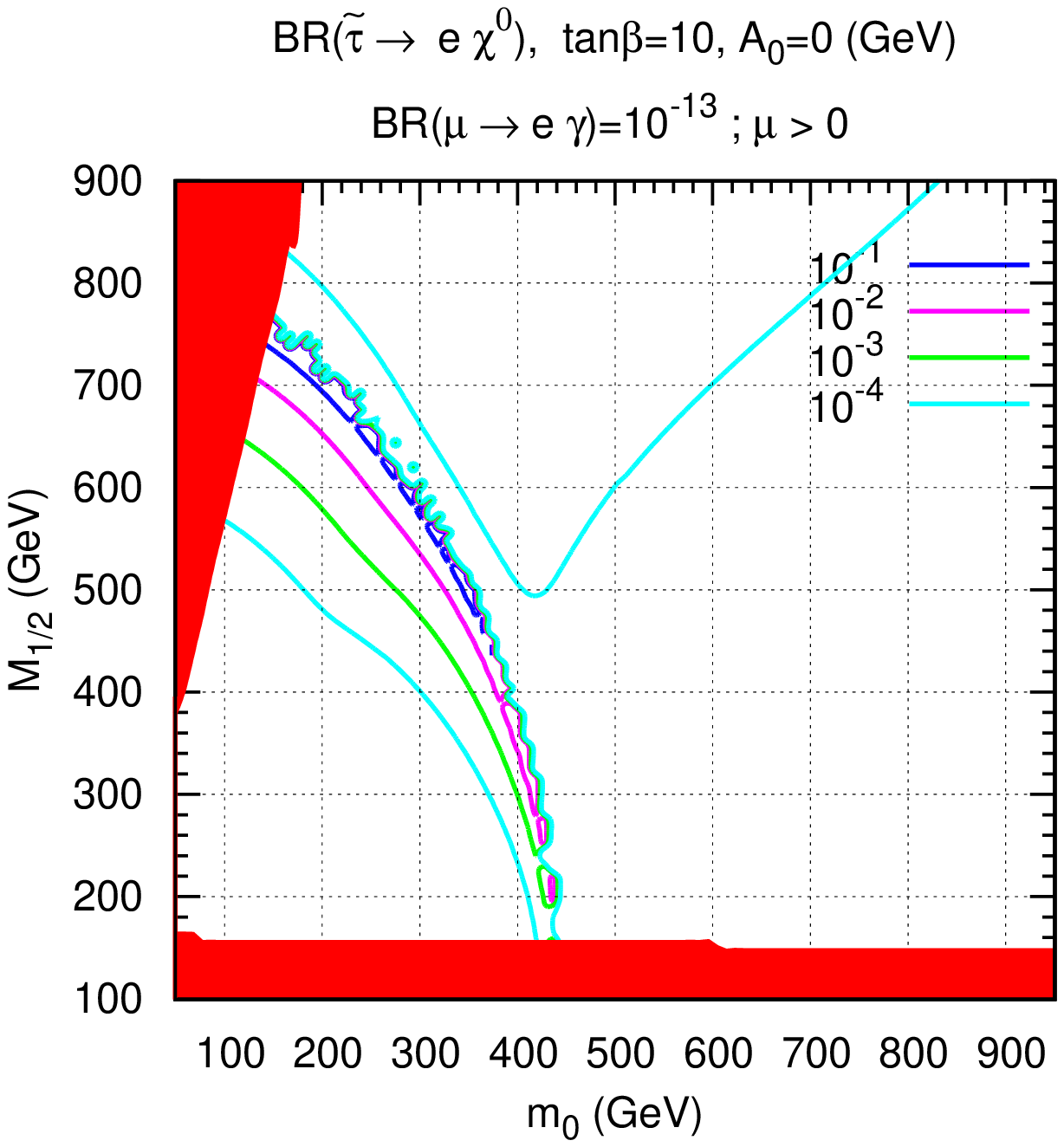}
  \end{tabular}
  \caption{Br(${\tilde\tau}_2 \to \mu +\chi^0_1$) (left panel) and
    Br(${\tilde\tau}_2 \to e +\chi^0_1$) (right panel), in the
    $m_0,M_{1/2}$ plane for our standard choice of parameters:
    $\mu>0$, $\tan\beta=10$ and $A_0=0$ GeV,  for type-I seesaw, imposing
    Br($\mu\to e +\gamma) \le 10^{-13}$.}
  \label{fig:4}
\end{figure}

The expected LFV branching ratios for ${\tilde\tau}_2 \to \mu
+\chi^0_1$ and ${\tilde\tau}_2 \to e +\chi^0_1$ depend on the
choice of the mSUGRA parameters. After a full scan over the mSUGRA
parameter space we found that the dependence on $A_0$ and on the sign
of $\mu$ is weaker, but that the rates decreased with
increasing values of $\tan\beta$. Therefore, we chose our standard
point with a relatively low value of $\tan\beta=10$, and for
definiteness took $\mu>0$, and $A_0=0$.  In Fig.~\ref{fig:3} we show
the contour plots for the LFV decays ${\tilde\tau}_2 \to \mu
+\chi^0_1$ (left panel) and ${\tilde\tau}_2 \to e +\chi^0_1$ (right
panel) in the $m_0,M_{1/2}$ plane for our standard choice of mSUGRA
parameters for the simplest pure type-I seesaw scheme. One sees
that there are regions in parameter space where the LFV decays of
the ${\tilde\tau}_2$ can be as large as of order $10^{-1}$. In
these plots the values of $M_R$ were chosen as to obtain the 
maximum LFV compatible with the present experimental limit of
Br($\mu\to e +\gamma) \le 1.2\cdot 10^{-11}$~\cite{pdg}.  Also shown
in these plots are the exclusion regions coming from the LEP
constraints on SUSY masses and also the exclusion obtained when
the neutralino is not the LSP \footnote{Note that we did not display
  the constraints coming from Dark Matter (DM) relic abundance.}. In
Fig.~\ref{fig:4} we show the same contour plots for Br($\mu\to e
+\gamma) \le 10^{-13}$, which will be achievable in the coming
experiments~\cite{MEG}.
Also in this case one observes in Fig.~\ref{fig:4} that the LFV
stau decay rates may exceed the 10\% level.
Notice also that the nontrivial features present in in
Figs.~\ref{fig:3} and Fig.~\ref{fig:4} reflect the well-known
cancellations between chargino and neutralino contributions to
$\mu\to e+\gamma$ already discussed above.

\begin{figure}[!htb]
  \centering
  \begin{tabular}{cc}
\includegraphics[width=0.48\textwidth]{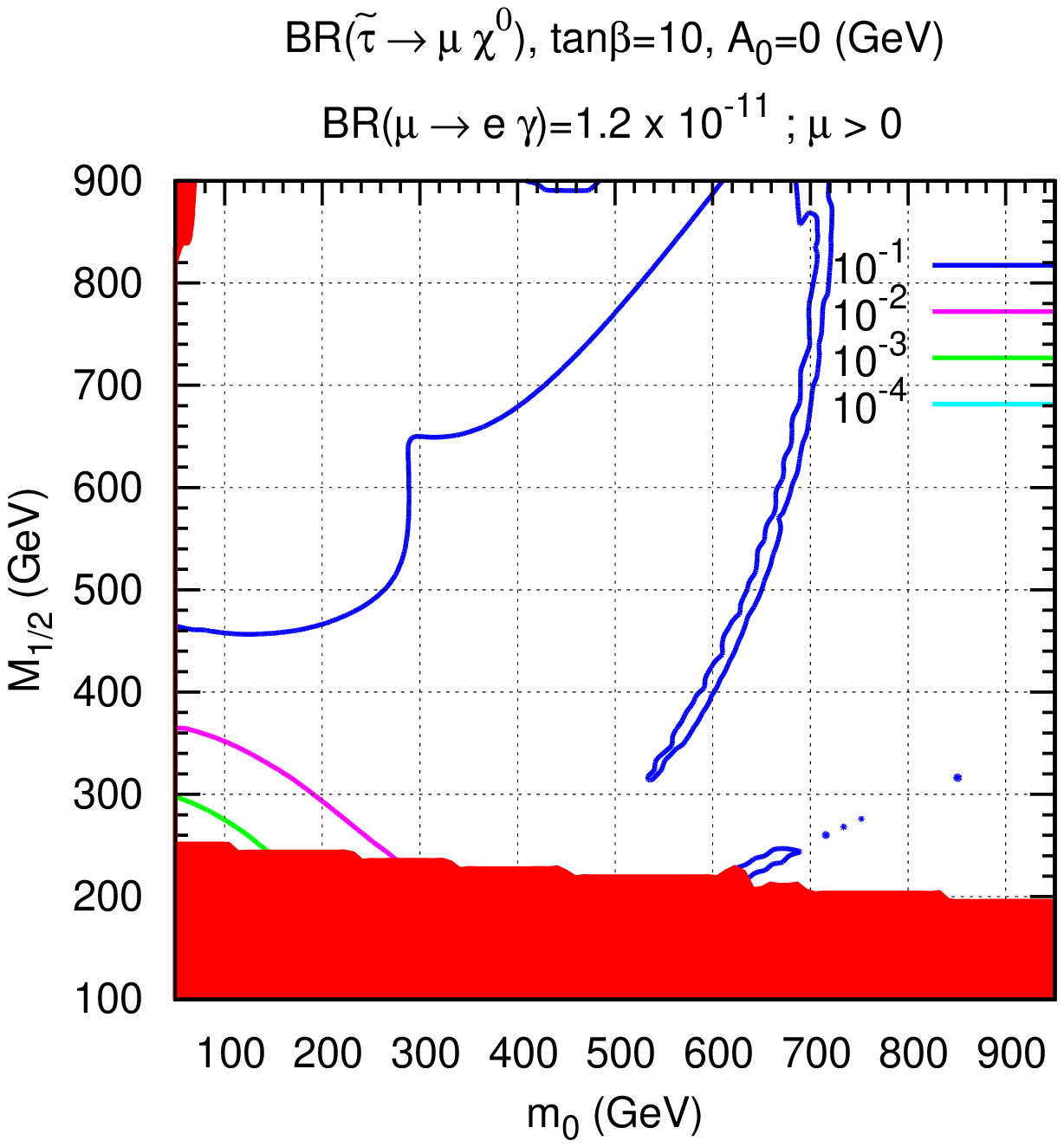}&
\includegraphics[width=0.48\textwidth]{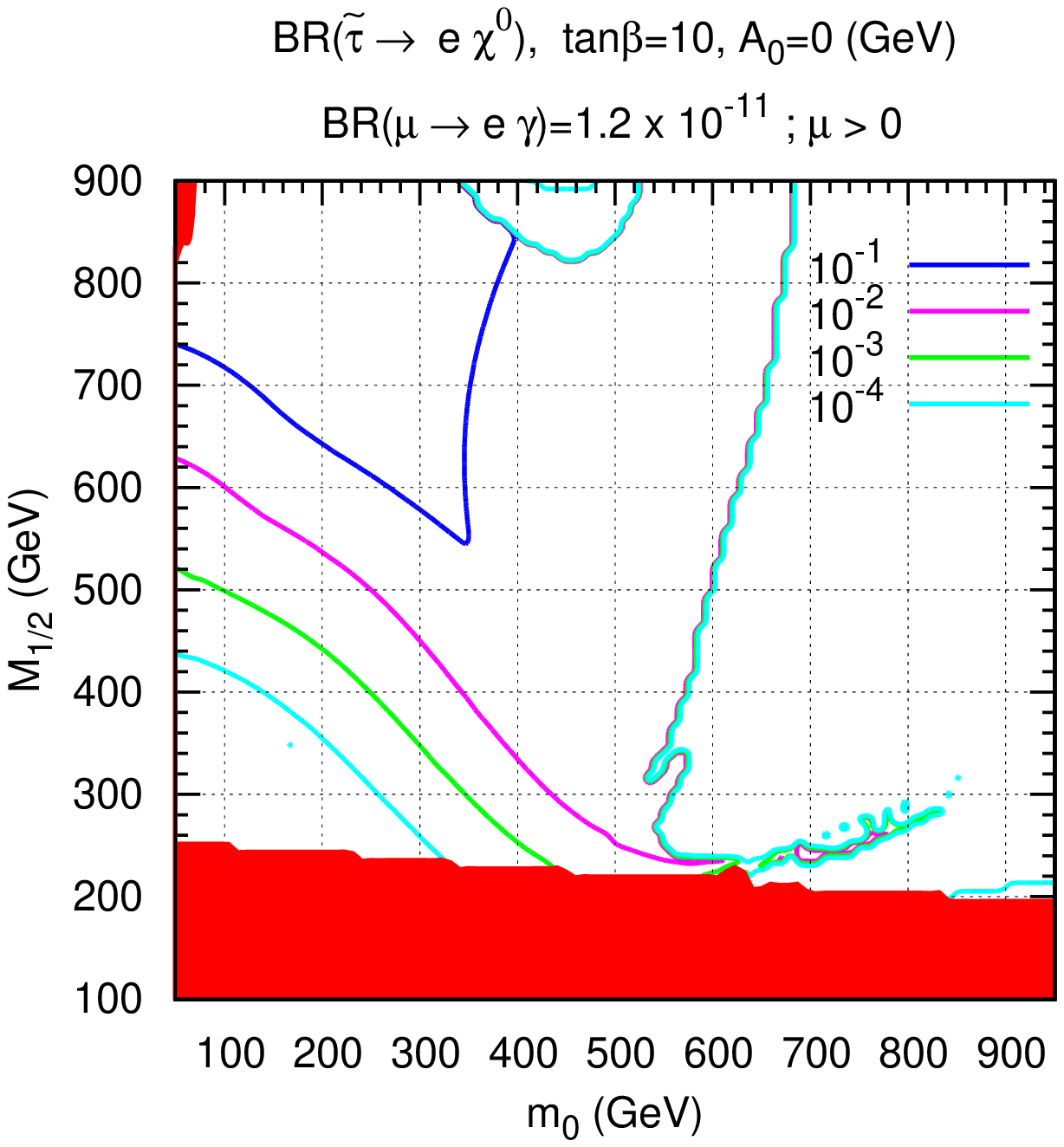}
  \end{tabular}
  \caption{Br(${\tilde\tau}_2 \to \mu +\chi^0_1$) (left panel) and
    Br(${\tilde\tau}_2 \to e +\chi^0_1$) (right panel), in the
    $m_0,M_{1/2}$ plane  for $\lambda_1=0.02$ and $\lambda_2=0.5$ and
    our standard choice of parameters: $\mu>0$, $\tan\beta=10$ and
    $A_0=0$ GeV, for type-II seesaw, imposing Br($\mu\to e +\gamma) \le 1.2\cdot
    10^{-11}$.} 
  \label{fig:5}
\end{figure}

\begin{figure}[!htb]
  \centering
  \begin{tabular}{cc}
\includegraphics[width=0.48\textwidth]{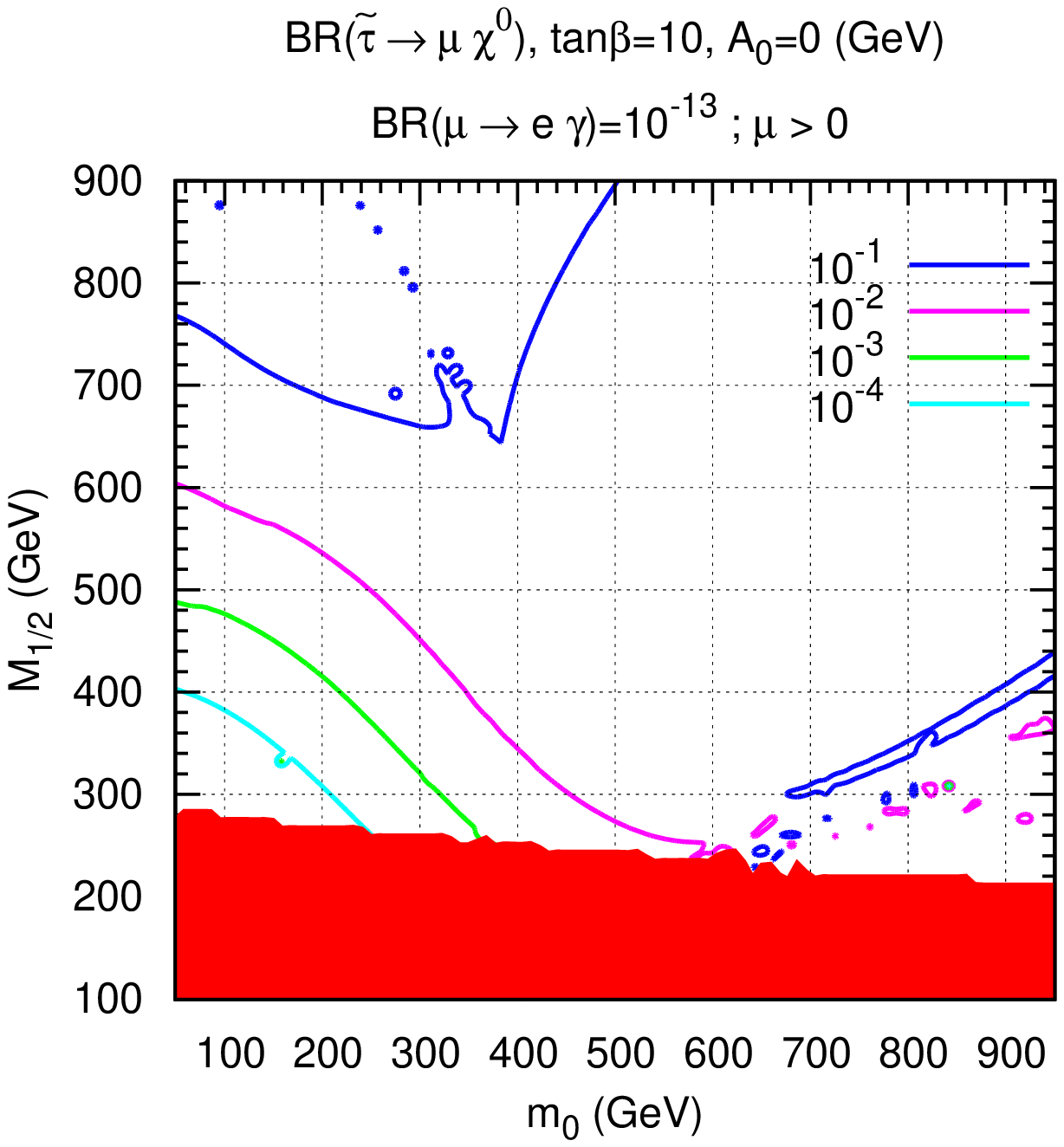}&
\includegraphics[width=0.48\textwidth]{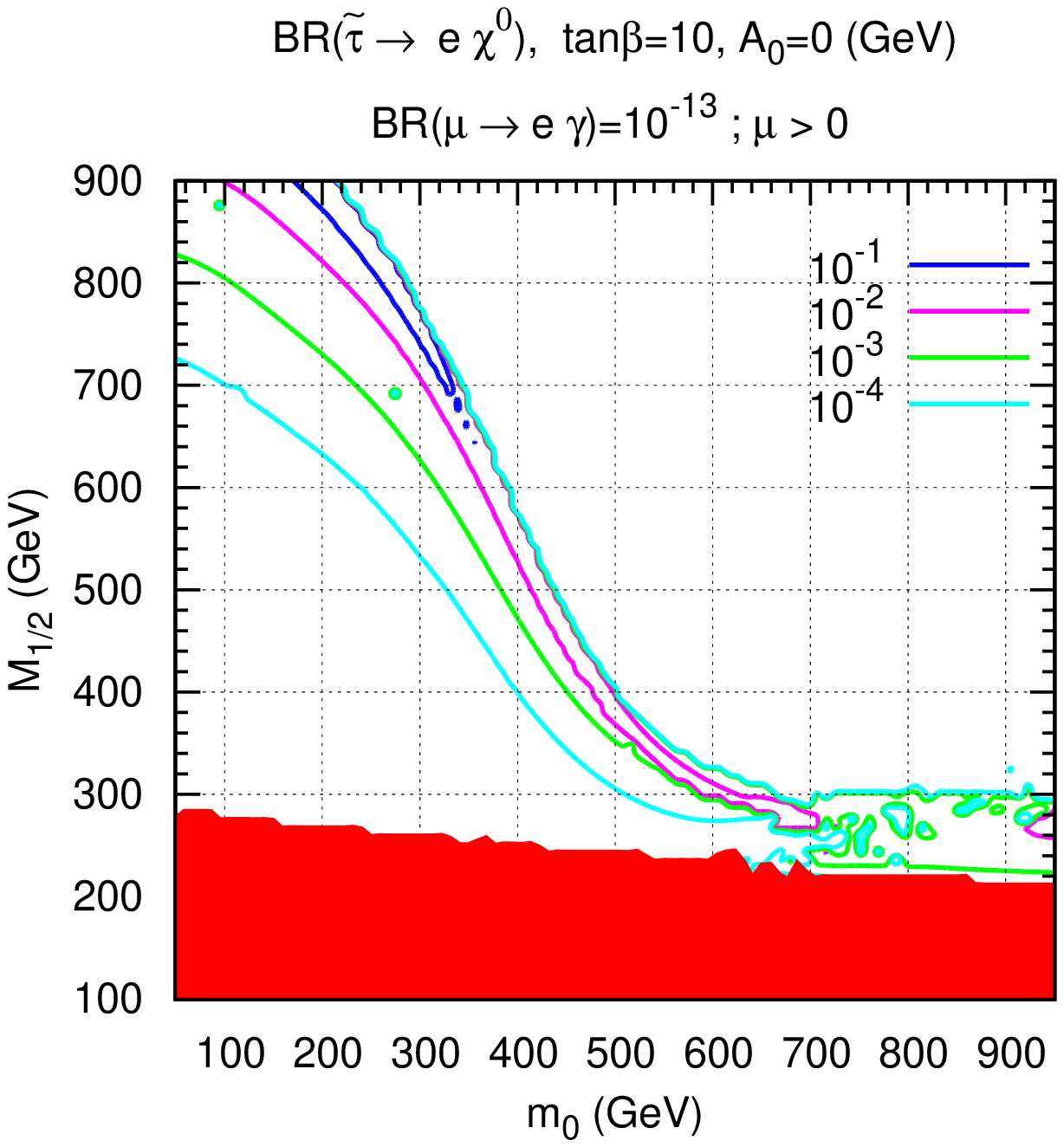}
  \end{tabular}
  \caption{Br(${\tilde\tau}_2 \to \mu +\chi^0_1$) (left panel) and
    Br(${\tilde\tau}_2 \to e +\chi^0_1$) (right panel), in the
    $m_0,M_{1/2}$ plane, for $\lambda_1=0.02$ and $\lambda_2=0.5$ and
    our standard choice of parameters: $\mu>0$, $\tan\beta=10$ and
    $A_0=0$ GeV, for type-II seesaw, imposing Br($\mu\to e +\gamma)
    \le 10^{-13}$.}
  \label{fig:6}
\end{figure}

\begin{figure}[!htb]
  \centering
  \begin{tabular}{cc}                    
  \includegraphics[width=0.48\textwidth]{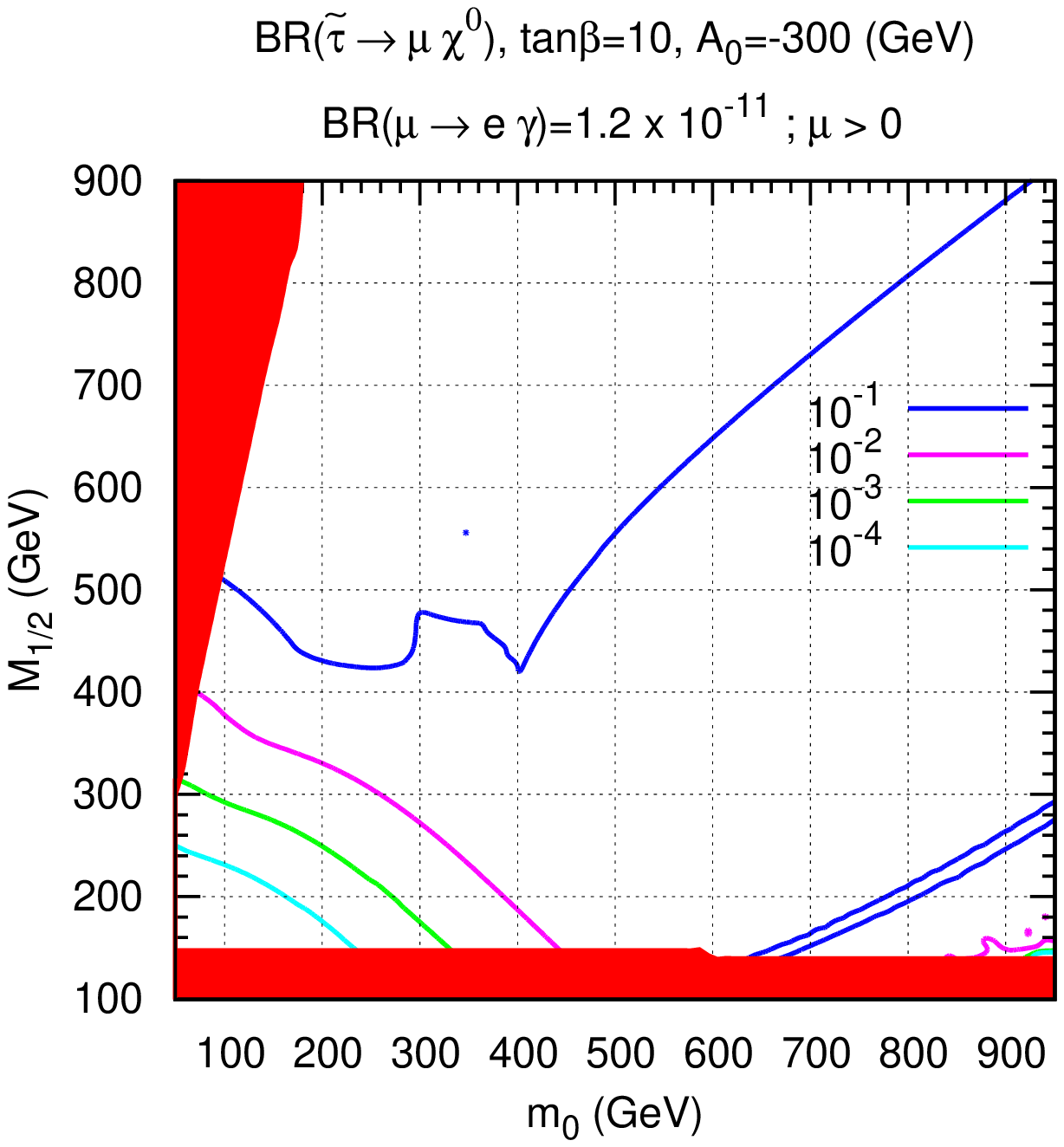}&
    \includegraphics[width=0.48\textwidth]{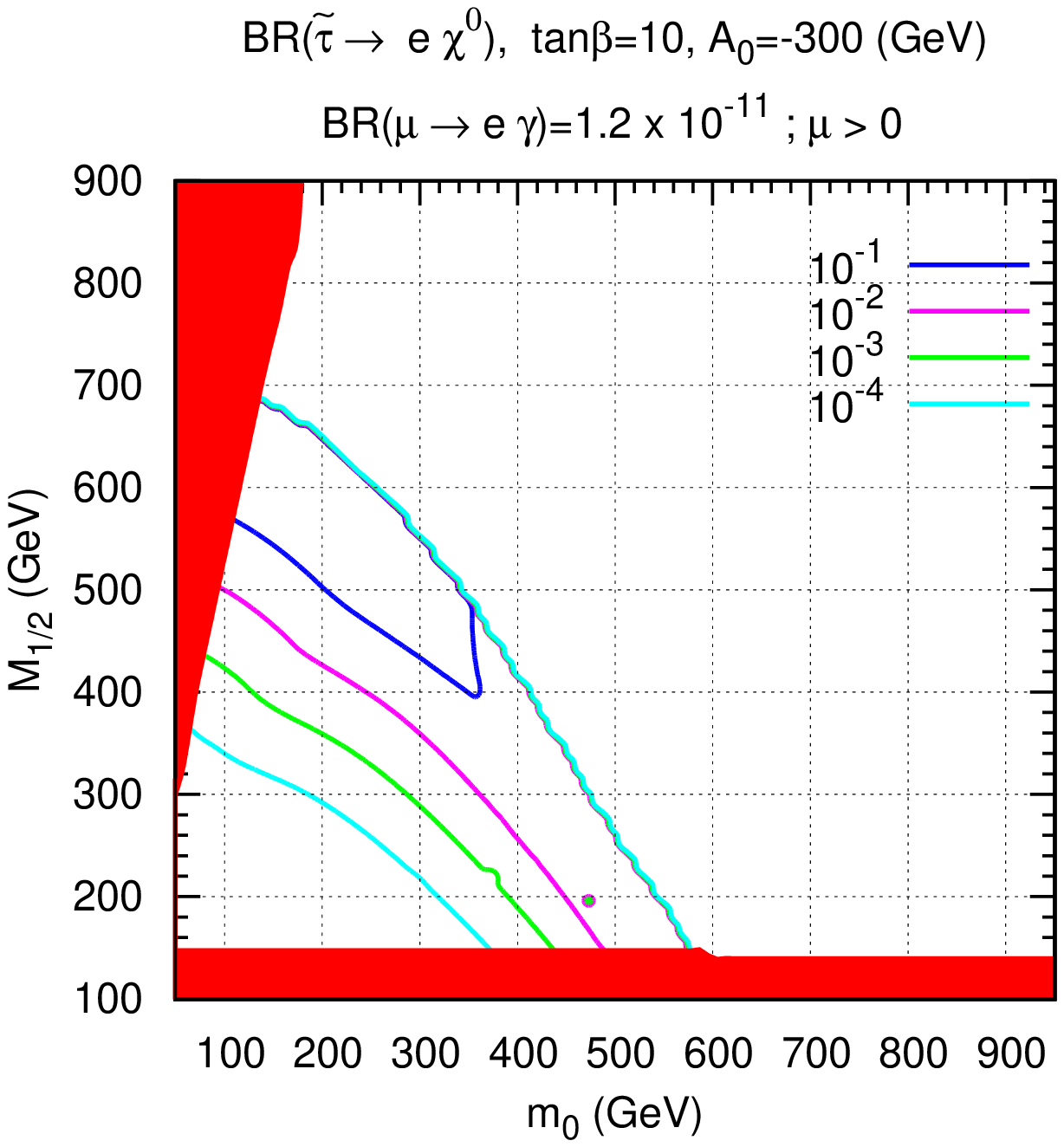}
  \end{tabular}
  \caption{Br(${\tilde\tau}_2 \to \mu +\chi^0_1$) (left panel) and
    Br(${\tilde\tau}_2 \to e +\chi^0_1$) (right panel), in the
    $m_0,M_{1/2}$ plane
    for standard choice of parameters: $\mu>0$, $\tan\beta=10$ but
    different $A_0=-300$ GeV, for type-I seesaw, imposing Br($\mu\to e
    +\gamma) \le 1.2\cdot 10^{-11}$.} 
  \label{fig:7}
\end{figure}

\begin{figure}[!htb]
  \centering
  \begin{tabular}{cc}
  \includegraphics[width=0.48\textwidth]{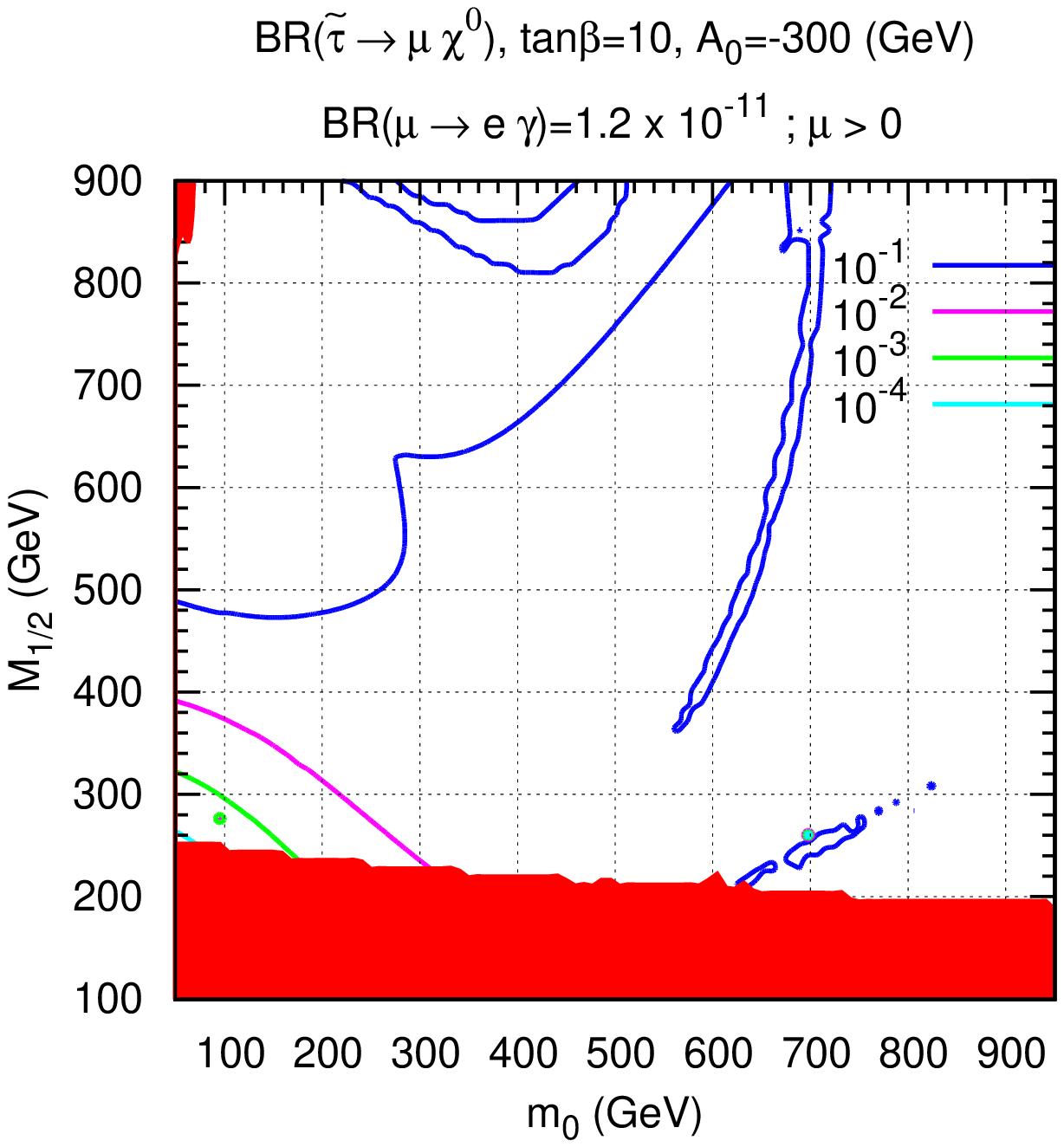}&
    \includegraphics[width=0.48\textwidth]{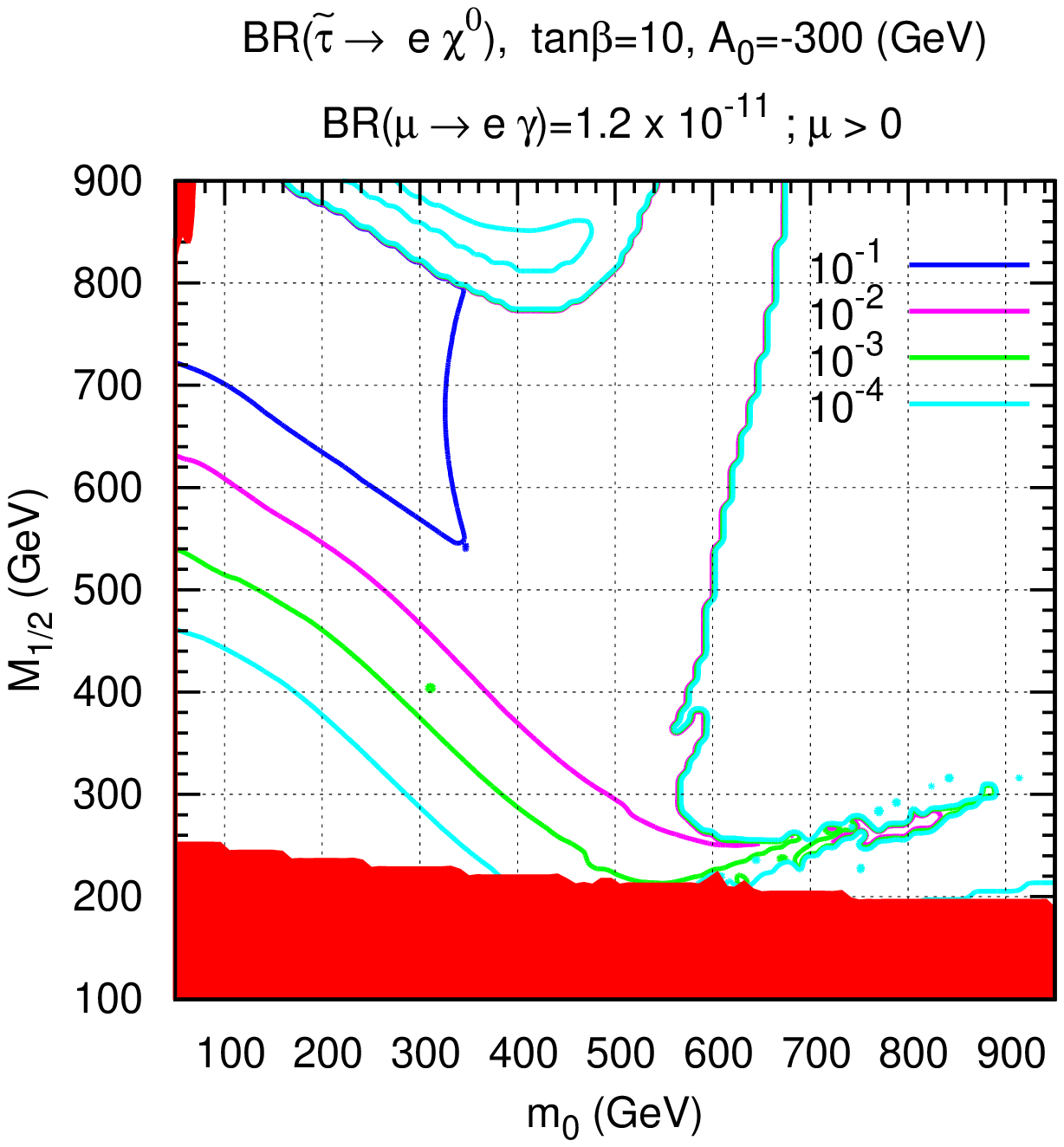}
  \end{tabular}
  \caption{Br(${\tilde\tau}_2 \to \mu +\chi^0_1$) (left panel) and
    Br(${\tilde\tau}_2 \to e +\chi^0_1$) (right panel), in the $m_0,M_{1/2}$ plane
    for $\lambda_1=0.02$ and $\lambda_2=0.5$ and
    standard choice of parameters: $\mu>0$, $\tan\beta=10$ but
    different $A_0=-300$ GeV, for type-II seesaw, imposing Br($\mu\to
    e +\gamma) \le 1.2\cdot 10^{-11}$.}
  \label{fig:8}
\end{figure}

\begin{figure}[!htb]
  \centering
  \begin{tabular}{cc}
  \includegraphics[width=0.48\textwidth]{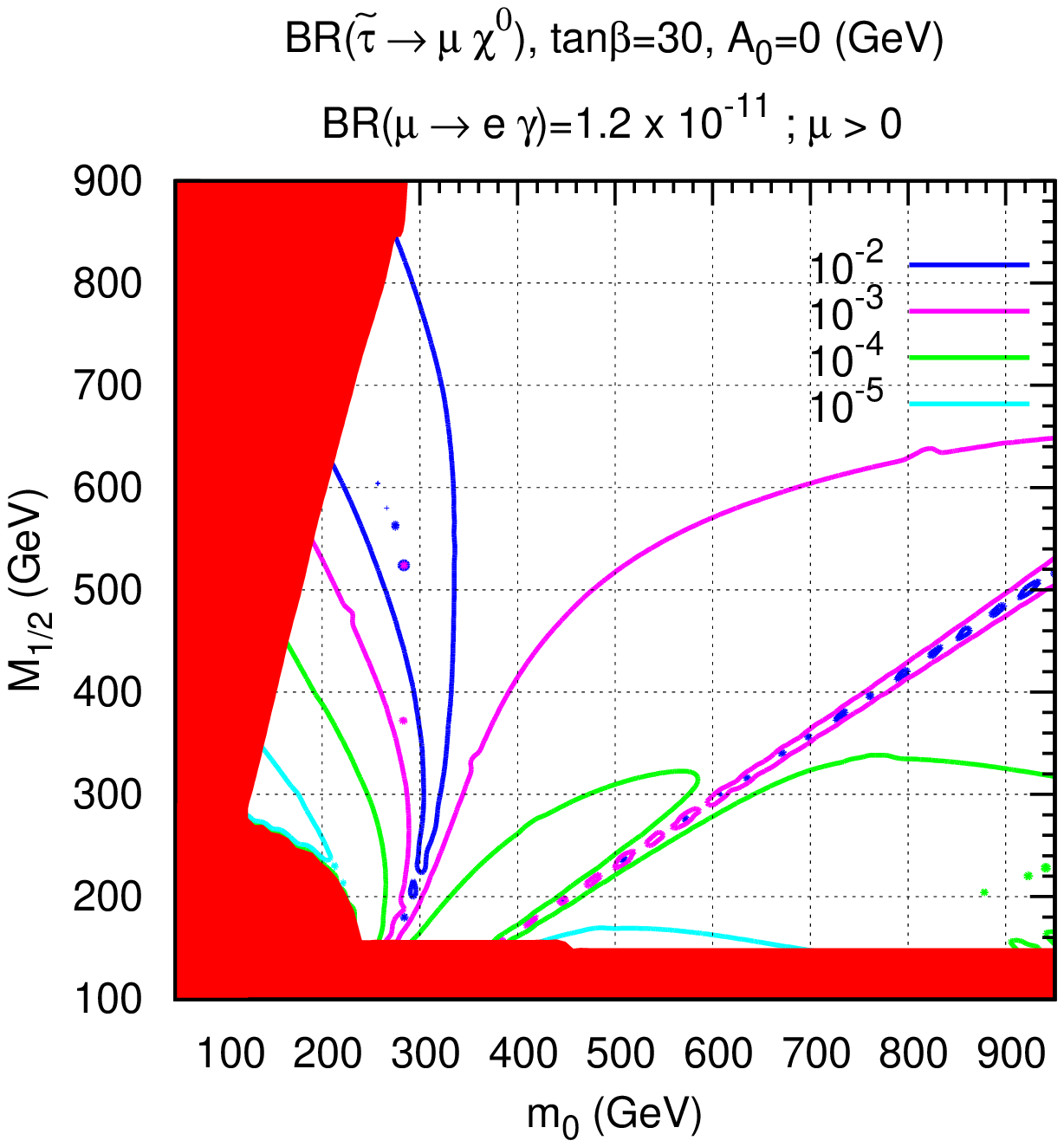}&
    \includegraphics[width=0.48\textwidth]{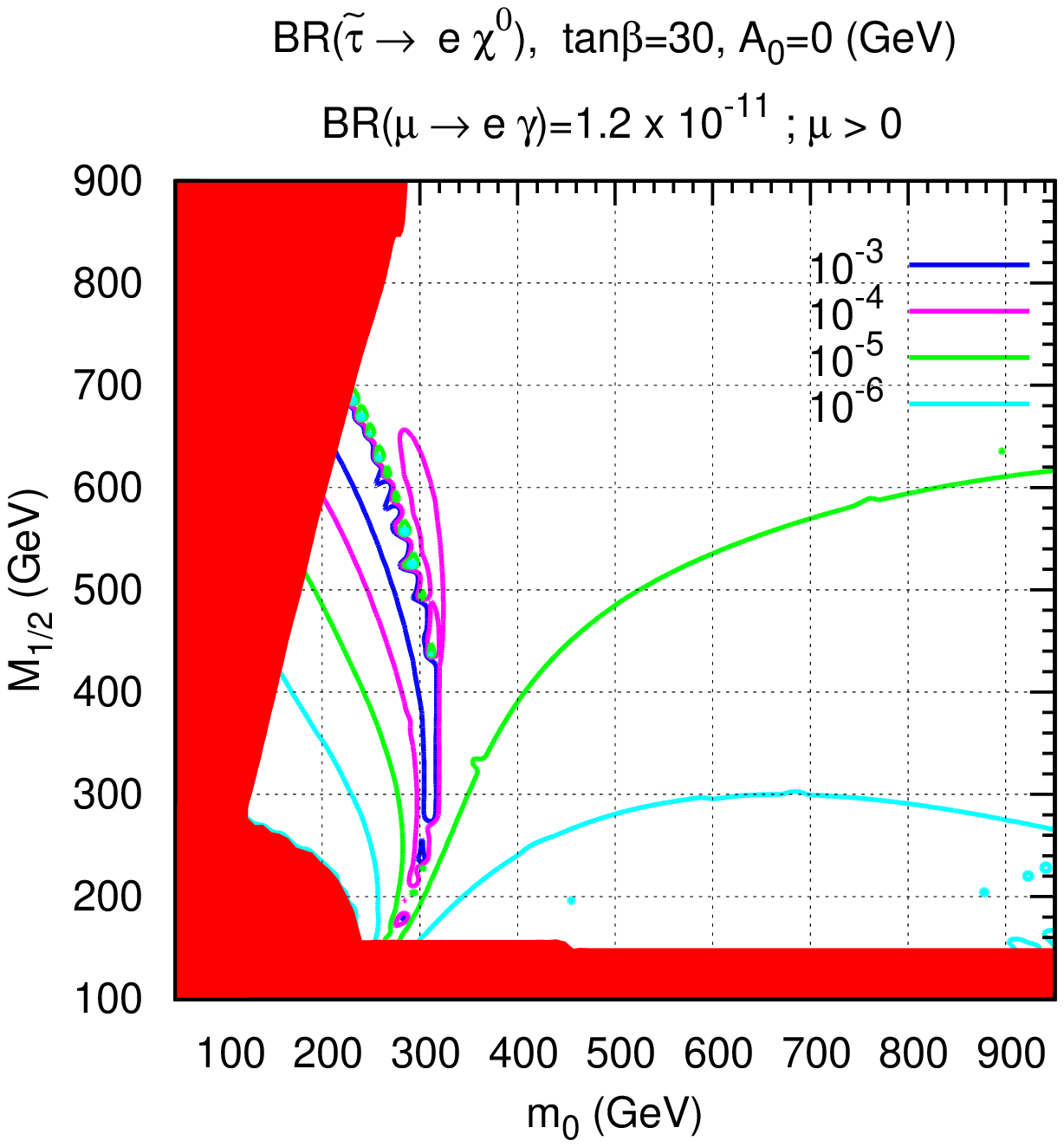}
  \end{tabular}
  \caption{Br(${\tilde\tau}_2 \to \mu +\chi^0_1$) (left panel) and
    Br(${\tilde\tau}_2 \to e +\chi^0_1$) (right panel), in the
    $m_0,M_{1/2}$ plane
    for standard choice of parameters: $\mu>0$, $A_0=0$ but
    different $\tan\beta=30$, for type-I seesaw, imposing Br($\mu\to e
    +\gamma) \le 1.2\cdot 10^{-11}$.} 
  \label{fig:9}
\end{figure}

\begin{figure}[!htb]
  \centering
  \begin{tabular}{cc}
  \includegraphics[width=0.48\textwidth]{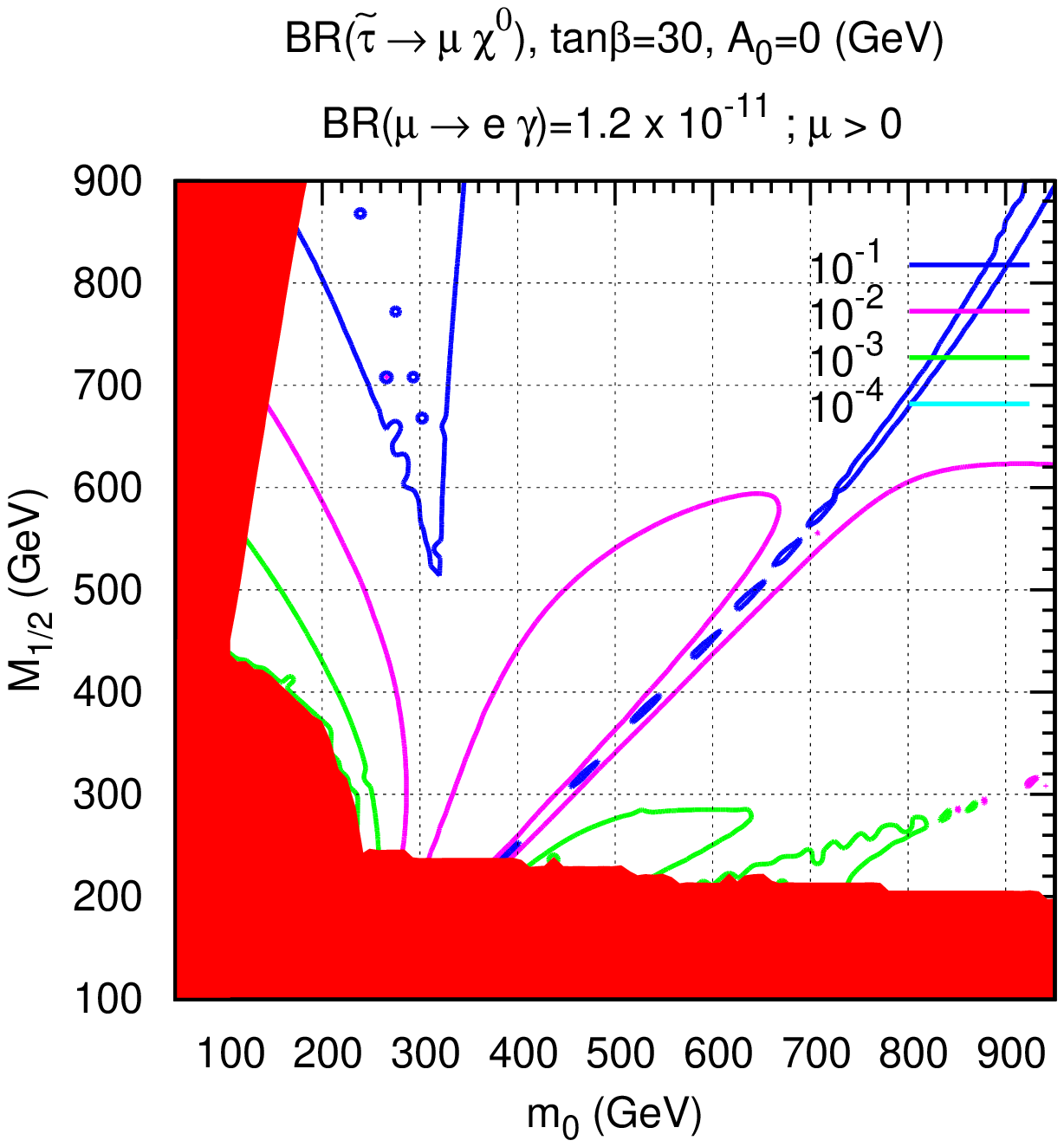}&
    \includegraphics[width=0.48\textwidth]{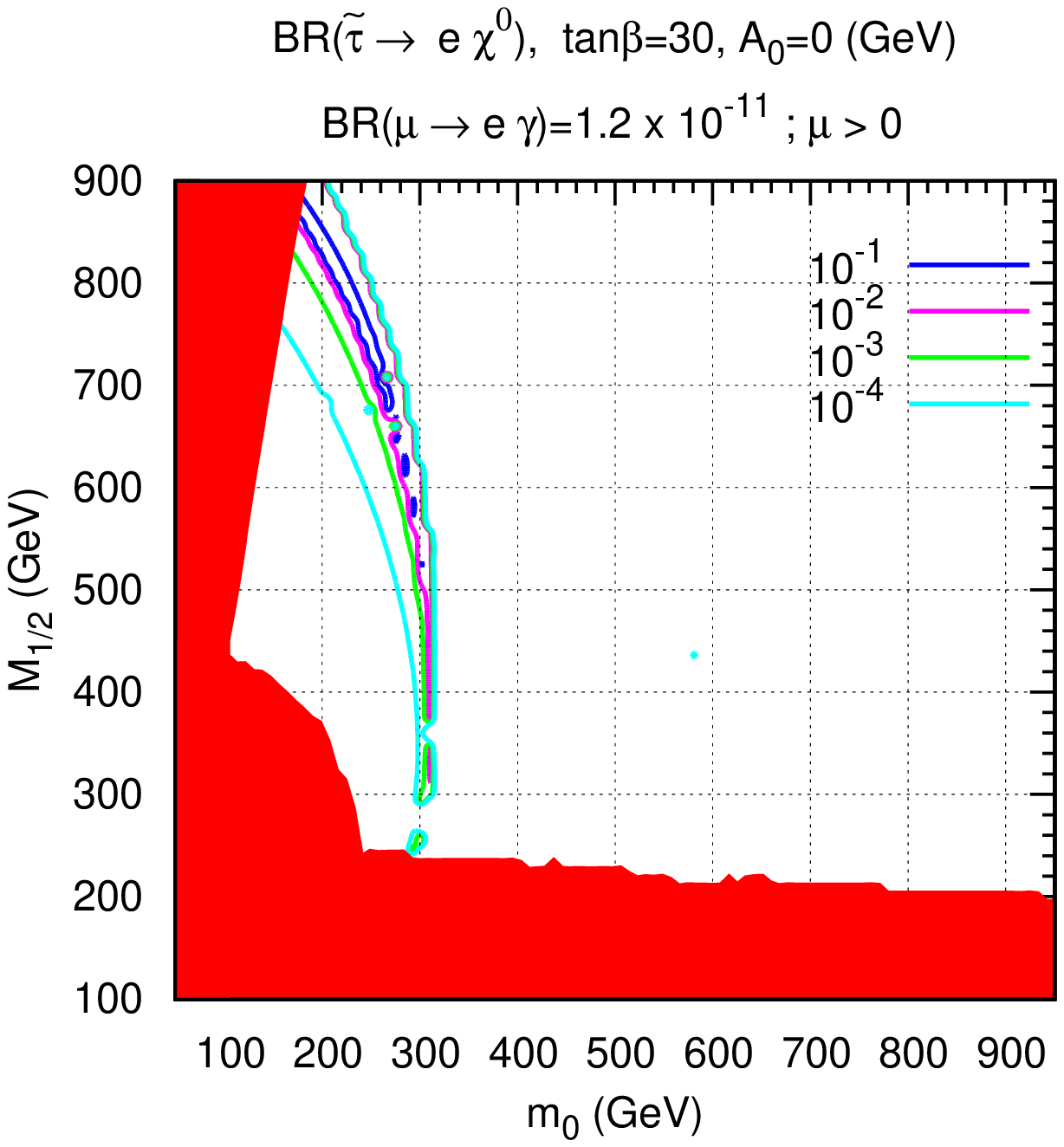}
  \end{tabular}
  \caption{Br(${\tilde\tau}_2 \to \mu +\chi^0_1$) (left panel) and
    Br(${\tilde\tau}_2 \to e +\chi^0_1$) (right panel), for
    $\lambda_1=0.02$ and $\lambda_2=0.5$, in the $m_0,M_{1/2}$ plane
    for standard choice of parameters: $\mu>0$, $A_0=0$, but different
    $\tan\beta=30$, for type-II seesaw, imposing Br($\mu\to e +\gamma)
    \le 1.2\cdot 10^{-11}$.}
  \label{fig:10}
\end{figure}

In Fig.~\ref{fig:5} and Fig.~\ref{fig:6} the same type of plots are
shown for type-II seesaw.  A comparison of these figures shows that,
qualitatively, the behavior is very similar for the two types of
seesaw. In both cases, the larger rates for ${\tilde\tau}_2 \to e
+\chi^0_1$ are more constrained in parameter space than those for
${\tilde\tau}_2 \to \mu +\chi^0_1$. Notice however that there
is an important difference between type-I and type-II seesaw, coming
from the presence of the Higgs triplets that contribute sizeably to
the running of the type-II beta functions. This gets reflected in
the supersymmetric particle spectra and hence in the shapes of the
red (shaded) regions in Fig.~\ref{fig:5} and Fig.~\ref{fig:6}.
One can observe, indeed, that the regions where the stau is the
lightest supersymmetric particle, as well as the regions already
excluded by LEP2 are substantially different for type-II seesaw, as
compared to the corresponding ones for type-I.  This follows from
the modification in the beta functions introduced by the addition of
the Higgs triplets, making $M_1$ and $M_2$ smaller in type-II than in 
type-I seesaw for the same value of $M_{1/2}$. 
The variation with the mSUGRA parameters is illustrated in
Fig.~\ref{fig:7} (type-I) and Fig.~\ref{fig:8} (type-II) for the
parameter $A_0$ and in Fig.~\ref{fig:9} (type-I) and Fig.~\ref{fig:10}
(type-II) for $\tan\beta$.  We can see that there is not much
variation with $A_0$, while the rates decrease rapidly with increasing
values of $\tan\beta$. The reason for this is that BR($\mu\to e
+\gamma$) increases as $\tan^4\beta$, thus constraining more
strongly the maximum attainable stau LFV rates. This effect is
stronger for type-I as can be seen by noting the different values for
the contour levels in Fig.~\ref{fig:9} and Fig.~\ref{fig:10}. The
variation with the sign of $\mu$ is weak and we do not show it
here. So, in summary, large LFV rates prefer moderate values of
$\tan\beta$ and this explains a posteriori the choice of our standard 
parameters.

\subsection{Total production cross section of $\chi_2^0$} 

As important as having a large branching ratio into a LFV
final state, is to be able to produce a large enough event 
sample. In order to estimate the number of LFV events expected at
the LHC, one notes that, from Figs.~\ref{fig:3}~-~\ref{fig:10}, in
the regions where the LFV is sizeable, the direct production of
staus at the LHC is negligible compared to that which arises from
cascade decays of heavier neutralinos, mainly $\chi_2^0$. 
We focus on the $\chi_2^0$, because decays such 
as $\chi_2^0 \to \mu\tau\chi^0_1$ are sensitive to flavour violation, 
whereas in the corresponding chargino decays the flavour information is 
lost.
Hence we first compute the total $\chi_2^0$ production cross
section.
\begin{figure}[!htb]
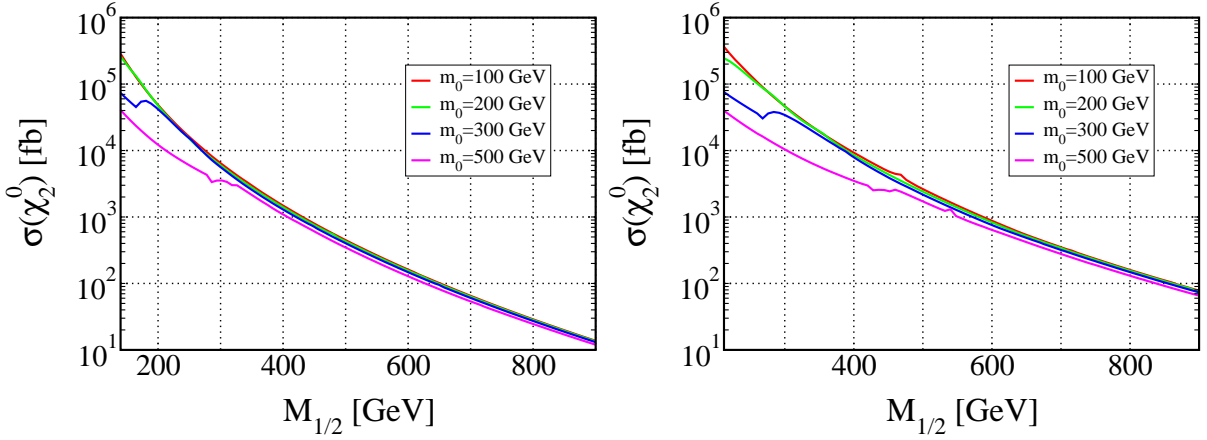

  \centering
  \includegraphics[width=0.48\textwidth]{plot-sigmaLOfb-m12_-_4m0.eps}
  \includegraphics[width=0.48\textwidth]{plot-sigmaLOfb-m12_-_4m0_-_II.eps}
  \caption{Production cross section (at leading order) of
    $\chi^0_2$ versus $M_{1/2}$ for varying $m_0$, and for our
    standard choice of parameters: $\mu>0$, $\tan\beta=10$ and $A_0=0$
    GeV, in type-I seesaw (left panel) and type-II seesaw (right
    panel) for  $\lambda_1=0.02$ and $\lambda_2=0.5$.  }
  \label{fig:Prod}
\end{figure}
In the left panel of Fig.~\ref{fig:Prod} we show the results for the
cross section for $\chi^0_2$ production as a function of $M_{1/2}$,
for different choices of $m_0$ and for our standard choice of
mSUGRA parameters: $\mu>0$, $\tan\beta=10$ and $A_0=0$ GeV, for the
pure type-I mSUGRA seesaw scheme. This choice of mSUGRA parameters
corresponds, as will be discussed below, to the case where the
branching ratios of the LFV stau decays are the largest.  This result was
obtained using the Prospino code~\cite{prospino} at Leading Order (LO)
approximation. We have checked that the Next to Leading Order (NLO)
calculation only changes the results slightly, due to an
appropriate choice of the renormalization scale~\cite{prospino}. So,
in all cross sections presented here, we only used the LO
approximation. 
The corresponding results for type-II seesaw are shown in the right
panel of  Fig.~\ref{fig:Prod}, for the same choice of mSUGRA parameters
and for $\lambda_1=0.02$ and $\lambda_2=0.5$. 

\subsection{Total production of $\chi_2^0$ times BR to $\mu$-$\tau$ lepton pair} 

In order to get an estimate of the expected number of LFV events
at the LHC we now use a combination of the Prospino and SPheno codes
to evaluate the product of the $\chi_2^0$ production cross
section times the branching ratios into LFV processes. Once we know
the luminosity at LHC we can multiply it with the above product 
to get the number of events.

\begin{figure}[!htb]
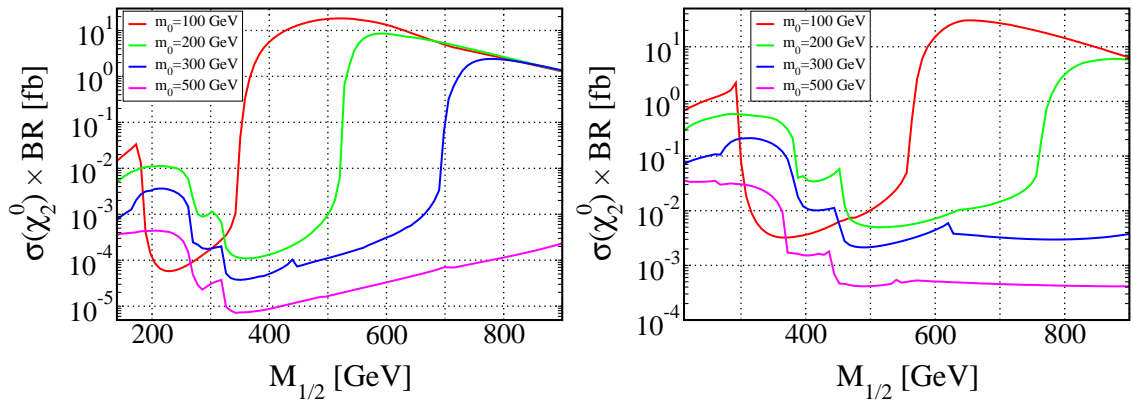

  \centering
\begin{tabular}{cc}
  \includegraphics[width=0.45\textwidth]{plot-sigmaLOfbBR-m12_-_4m0.eps} &
  \includegraphics[width=0.45\textwidth]{plot-sigmaLOfbBR-m12_-_4m0_-_II.eps}
\end{tabular}
\caption{Production cross section (at leading order) of $\chi^0_2$ times BR 
    of $\chi^0_2$ going to $\mu$-$\tau$ lepton pair versus $M_{1/2}$
    for $m_0=100$~GeV (red), 200~GeV (green), 300~GeV (blue) and
    500~GeV (magenta), and for our standard choice of  
    parameters: $\mu>0$, $\tan\beta=10$ and $A_0=0$ GeV, for type-I
    (left panel) and for type-II seesaw (right panel) with  $\lambda_1=0.02$ and
    $\lambda_2=0.5$, imposing Br($\mu\to e +\gamma) \le
    1.2\cdot 10^{-11}$.}  
  \label{fig:ProdXBR}
\end{figure}

In Fig.~\ref{fig:ProdXBR}, we have plotted, for type-I seesaw (left
panel) and type-II (right panel), the production cross section at
leading order of the second lightest neutralino $\sigma(\chi^0_2)$
times the BR of $\chi^0_2$ going to the opposite-sign dilepton signal
$\chi^0_1\,\mu\,\tau$ as a function of $M_{1/2}$, for different values
of $m_0$. We have fixed the rest of the mSUGRA parameters to our
standard mSUGRA point and imposed an upper limit on Br($\mu\to e
+\gamma) \le 1.2\cdot 10^{-11}$.  In type-I seesaw, the number of
events of the opposite-sign dilepton signal
$\chi^0_2\to\chi^0_1\,\mu\,\tau$ can be of the order of $10^3$ for
$m_0\sim 100$ GeV and $M_{1/2}\sim[450,\,600]$ GeV, assuming a
luminosity ${\cal L} = 100\ \mathrm{fb}^{-1}$.  In type-II seesaw,
there can be a maximum number of events of the order of $10^3$ for
$m_0\sim 100$ GeV and $M_{1/2}\sim[600,\, 800]$ GeV.

\begin{figure}[!htb]
  \centering
  \includegraphics[width=0.45\textwidth]{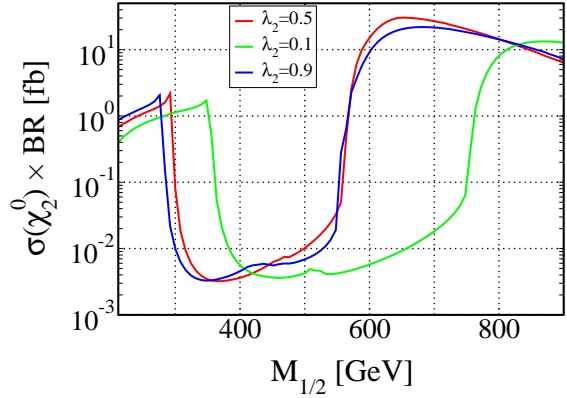}
  \caption{
   Production cross section (at leading order) of $\chi^0_2$ times BR 
    of $\chi^0_2$ going to $\mu$-$\tau$ lepton pair versus $M_{1/2}$, 
    for our standard choice of  
    parameters: $\mu>0$, $\tan\beta=10$ and $A_0=0$ GeV, for type-II
    seesaw, imposing Br($\mu\to e +\gamma) \le 1.2\cdot 10^{-11}$, 
    for a fixed value of $m_0=100$ GeV and different values of 
    $\lambda_2=0.1$~(green), 0.5~(red), 0.9~(blue).}
  \label{fig:ProdXBR2}
\end{figure}

For type-II seesaw where we have less parameters, we can look at
variations of the result with the values of the triplet Higgs boson
coupling $\lambda_2$, a parameter that can not be determined from
neutrino data alone as it appears only in the ratio $\lambda_2/M_T$,
see eq.~(\ref{eq:ssII}). In Fig.~\ref{fig:ProdXBR2} we show the
dependence of the product of cross section times LFV branching ratios
as function of $\lambda_2$ for our standard point. We should mention
that the other Higgs boson triplet coupling $\lambda_1$, does not
contribute to LFV decays, and hence is left undetermined by this
analysis.

As has been discussed in \cite{Carquin:2008gv}, the dominant 
standard model backgrounds for the process considered are expected to 
be $WW$ and $t{\bar t}$ production. The cuts necessary to reduce this 
background will depend on the details of the SUSY spectrum and a 
detailed investigation is beyond the scope of this paper. The 
results of \cite{Carquin:2008gv} suggest that the signal should 
be visible for $\sigma(\chi^0_2)\times$BR of order 
${\cal O}(10)$ fb.

\section{Conclusions and outlook}
\label{sec:cncl}

Low energy neutrino experiments, including oscillation studies and
neutrinoless double-beta decay searches may, optimistically, determine
at most 9 neutrino parameters: the 3 neutrino masses, the 3 mixing
angles and potentially the 3 CP violating phases.
This is insufficient to fully reconstruct the underlying mechanism of
neutrino mass generation.
Under the assumption that neutrino masses arise \emph{a la seesaw}, we
have considered the simplest pure type-I or pure type-II 
seesaw schemes in mSUGRA. 

We have performed a full scan over the mSUGRA parameter
space in order to identify regions where LFV decays of $\chi^0_2$ can
be maximal, while still respecting low-energy constraints that follow
from the upper bounds on Br($\mu\to e \gamma$).  We have also
estimated the expected number of events for $\chi^0_2\to \chi^0_1 +
\tau +\mu $, for a sample luminosity of ${\cal L} = 100\
\mathrm{fb}^{-1}$.  The expected number of events for the other
channel $\chi^0_2\to \chi^0_1 + \tau + e $ is always smaller, as can
be seen from the LVF branching ratios presented in
section~\ref{sec:staudecays}.
We have found that the pure seesaw-II scheme is substantially simpler
and comes closer to be fully reconstructable, provided
additional LFV decays are detected and some supersymmetric particles
are discovered at the Large Hadron Collider.

Note that in what concerns the expected maximum number of events both type-I
and type-II schemes give similar results. However, as we have seen,
given their smaller number of parameters, type-II seesaw schemes are
more likely to be reconstructable through a combination of low
energy neutrino measurements, with the possible detection of
supersymmetric states and lepton flavour violation at the LHC.
This should encourage one to perform full-fledged
dedicated simulations, in order to ascertain their feasibility
within realistic experimental conditions~\cite{delAguila:2008iz}.

Finally we note that we have not exploited the fact the LFV might 
induce new ``edge variables'', giving additional information 
\cite{Bartl:2005yy}. We have focused here on LHC, but mention 
that a future ILC would be much more suited for measuring LFV 
SUSY processes \cite{Krasnikov:1995qq,ArkaniHamed:1996au,ArkaniHamed:1997km,%
Hisano:1998wn,Nomura:2000zb,Porod:2002zy,Deppisch:2003wt}.

\section*{Acknowledgements}

Work supported by Spanish grants FPA2008-00319/FPA and Accion
Integrada HA-2007-0090 (MEC).  The work of A.V.M. is supported by {\it
  Funda\c c\~ao para a Ci\^encia e a Tecnologia} under the grant
SFRH/BPD/30450/2006. The work of J.C.R. and A.V.M is also supported by
the RTN Network MRTN-CT-2006-035505 and by {\it
  Funda\c c\~ao para a Ci\^encia e a Tecnologia} through the projects
CFTP-FCT UNIT 777,  POCI/81919/2007 and  CERN/FP/83503/2008.
W.P.~is partially supported by the German
Ministry of Education and Research (BMBF) under contract 05HT6WWA, by
the DAAD, project number D/07/13468.

\end{document}